\documentclass[10pt]{article}

\newlength{\minitwocolumn}
\font\teneufm=eufm10
\font\seveneufm=eufm7
\font\fiveeufm=eufm5
\newfam\eufmfam
\textfont\eufmfam=\teneufm
\scriptfont\eufmfam=\seveneufm
\scriptscriptfont\eufmfam=\fiveeufm

\makeatletter
\@addtoreset{equation}{section}
\makeatother

\title{\bf
\Large{\bf 
Form factors of the half-infinite
XXZ spin chain \\
with a triangular boundary}}
\begin{document}

\maketitle

\begin{center}
{P. Baseilhac$~^{\alpha}$} and {T. Kojima$~^{\beta}$}
\\~\\
{\it {$\alpha$}~
Laboratoire de Math\'ematiques et Physique Th\'eorique CNRS/UMR 7350,
\\F\'ed\'eration Denis Poisson FR2964,
Universit\'e de Tours,
Parc de Grammont, 37200 Tours, 
FRANCE\\
baseilha@lmpt.univ-tours.fr
\\
~
\\
{$\beta$}~
Department of Mathematics and Physics,
Faculty of Engineering,
Yamagata University,\\
 Jonan 4-3-16, Yonezawa 992-8510, JAPAN\\
kojima@yz.yamagata-u.ac.jp}
\end{center}

~\\

\begin{abstract}
The half-infinite XXZ spin chain 
with a triangular boundary
is considered in the massive regime.
Two integral representations 
of form factors of local operators are proposed using bosonization. 
Sufficient conditions such that the expressions for triangular boundary conditions coincide with 
those for diagonal boundary conditions are identified. 
The expressions are compared with known results upon specializations.
Using the spin-reversal property which relates the Hamiltonian with upper and lower triangular boundary conditions, new identities between multiple integrals of infinite products are extracted.
\end{abstract}

\section{Introduction}
Among the family of quantum integrable models, the XXZ spin chain with different types of integrable boundary conditions is the subject of current investigations. For periodic boundary conditions, several results are already known: the spectral problem - eigenvalues and eigenvectors - for the Hamiltonian has been handled either using the Bethe ansatz (BA) approach \cite{Bethe}, or using the corner transfer matrix method (CTM) in the thermodynamic limit \cite{Baxter}. 
In addition, correlation functions and form  factors of local operators have 
been derived using the quantum inverse scattering method (QISM) 
in \cite{KBI,KMT}, 
or using the vertex operator approach (VOA) 
in the thermodynamic limit
\footnote{For models associated with higher rank affine Lie algebras, 
the calculation of correlation functions and form factors using the QISM 
is currently investigated (see e.g. \cite{Reshetikhin,Wheeler}). 
In the thermodynamic limit, 
correlation functions can be derived using the 
VOA \cite{FJ,Koyama,JKK,Bernard,KSU}.} 
\cite{DFJMN}.
It has been checked that
expressions of correlation functions
obtained by both approaches coincide
in the thermodynamic limit. 
They are given by multiple integrals of meromorphic functions.

For general integrable boundary conditions, 
less is known. For diagonal \cite{Sklyanin}, triangular \cite{PL}, related non-diagonal boundaries \cite{Nepomechie,CLSW} or $q$ a root of unity \cite{MNS},  the spectral problem  of the corresponding finite XXZ open chain has been solved using the BA. Extending the analysis of \cite{KMT}, 
correlation functions have been also proposed for diagonal boundary conditions \cite{KZMNST,KZMNST2}. In the thermodynamic limit, the half-infinite XXZ spin chain with general integrable boundary conditions can be alternatively formulated using the VOA, either starting from Sklyanin's framework \cite{Sklyanin} or from the recently developed Onsager's approach \cite{BK,BB1} (see also related works \cite{currents}). For diagonal and triangular boundary conditions, eigenvalues and eigenvectors are proposed in \cite{JKKKM,BB1} and integral representations of correlation functions in the massive regime are computed in \cite{JKKKM,BK1}, respectively. In the massless regime and diagonal boundary conditions, note that integral representations are obtained in \cite{Kojima3}.

 Up to now, the calculation of form factors of local operators of the half-infinite XXZ spin chain within the VOA - even in the case of diagonal boundary conditions - has not been addressed. Besides the importance of the form factors in the non-perturbative analysis of this model, there are four main motivations for considering this problem in details. First, 
an alternative derivation of the QISM proposal for diagonal boundary conditions \cite{KZMNST,KZMNST2} is highly desirable. Secondly,  for {\it non-diagonal} boundary conditions there are no known results available yet for the form factors, besides the two integral representations of correlation functions recently proposed in \cite{BK1}.
Third, correlation functions and form factors are known to satisfy quantum boundary
Knizhnik-Zamolodchikov equations, whose solutions for non-diagonal boundary conditions are recently investigated in \cite{Ka,Stok}. Any new piece of information - for instance integral representations of form factors - is welcome, especially if it is related with\footnote{Type I and type II $q$-vertex operators entering in the correlation functions and form factors are intertwiners of infinite dimensional (level one) representations of the $q$-Onsager algebra, a coideal subalgebra of $U_q(\widehat{sl_2})$ \cite{BB1}.} the $q$-Onsager algebra 
representation theory.  Fourth, it is expected that the remarkable identities between the correlation functions given in the form of $n$-fold multiple integrals of elliptic theta functions and discovered in \cite{BK1} can be generalized to arbitrary form factors.

In this paper, the half-infinite XXZ spin chain 
with a triangular boundary
is studied in the massive regime.
This paper is a continuation of the paper \cite{BK1}.
Integral representations 
of form factors of local operators are proposed using bosonization. 
Using the spin-reversal property which relates the Hamiltonian with upper 
and lower triangular boundary conditions, new identities between 
multiple integrals of infinite products are extracted.
The paper is organized as follows.
In Section 2, we briefly recall how the half-infinite XXZ spin chain is formulated within the VOA: eigenvalues, eigenvectors, dual eigenvectors and scalar products between these are given. 
Integral representations of form factors of 
local operators for the half-infinite XXZ spin chain with 
{\it diagonal} and {\it triangular} boundary conditions are proposed
in Section 3 and Section 4, respectively.
In Section 3,
upon specializations the expressions are compared with 
the known results obtained within the QISM \cite{KZMNST} and \cite{JKKKM}. 
Some checks are done in simple cases. 
Included in Section 3 and Section 4,
using the spin-reversal property,
relations between multiple integrals of infinite products are extracted.
In Section 5 we summarize concluding remarks.
In Appendix \ref{appendix:A}, basic definitions of the $R$-matrix and the $K$-matrix are recalled.
In Appendix \ref{appendix:B} 
we summarize bosonizations of the vertex operators
and normal orderings.
In Appendix \ref{appendix:C} 
we give a direct proof of the spin reversal property for the
simplest case. 
In Appendix \ref{appendix:D}
we summarize convenient formulae for the calculations of 
vacuum expectation values.

\section{The half-infinite XXZ chain}
In this Section, 
basic necessary ingredients for the analysis of the form factors of local operators are briefly recalled. We refer the reader to \cite{JKKKM,BB1,BK1} for more details. 

\subsection{Hamiltonian}

In the present paper, we consider the half-infinite XXZ spin chain with Hamiltonian\footnote{We use the standard Pauli matrices
\begin{eqnarray}
\sigma^x=\left(\begin{array}{cc}
0&1\\
1&0
\end{array}\right),~
\sigma^y=
\left(\begin{array}{cc}
0&-\sqrt{-1}\\
\sqrt{-1}&0
\end{array}\right),~
\sigma^z=
\left(\begin{array}{cc}
1&0\\
0&-1
\end{array}\right),~
\sigma^+=\left(\begin{array}{cc}
0&1\\
0&0
\end{array}\right),~
\sigma^-=\left(\begin{array}{cc}
0&0\\
1&0
\end{array}\right).
\end{eqnarray} 
}:
\begin{eqnarray}
H_B^{(\pm)}=-\frac{1}{2}
\sum_{k=1}^\infty (\sigma_{k+1}^x \sigma_k^x+\sigma_{k+1}^y \sigma_k^y+\Delta \sigma_{k+1}^z \sigma_k^z)
-\frac{1-q^2}{4q}\frac{1+r}{1-r}\sigma_1^z-\frac{s}{1-r} \sigma_1^\pm .
\label{def:Hamiltonian}
\end{eqnarray}
Here, as usual $\Delta=(q+q^{-1})/2$ denotes the anisotropy parameter. Diagonal boundary conditions correspond to the special case $s=0$, whereas $s\neq 0$ describes  triangular boundary conditions of upper/lower type according to the sign index $(\pm)$ chosen in (\ref{def:Hamiltonian}). The massive regime corresponds to the range of parameters:
\begin{eqnarray}
~~~-1<q<0,~~~-1\leq r \leq 1,~~~s \in {\bf R}.
\end{eqnarray}
Note that under conjugation of $H_B^{(\pm)}$ by the spin-reversal operator 
$\hat{\nu}=\prod_{j=1}^\infty \sigma_j^x$ the sign of the boundary term is reversed. Then, in the discussion below we can restrict our attention to the domain where the boundary term $-\frac{1-q^2}{4q}\frac{1+r}{1-r}
\geq 0$, namely $-1 \leq r \leq 1$.

\subsection{The transfer matrix  in the vertex operator approach}
We now turn to the mathematical formulation of the half-infinite XXZ spin chain based on the $q$-vertex operator approach \cite{JKKKM,BB1}.
In Skylanin's framework the transfer matrix associated with the {\it finite} XXZ open spin chain 
is built from a finite number of $R$-matrices, and two $K$-matrices \cite{Sklyanin}. In order to formulate the {\it infinite} model 
(\ref{def:Hamiltonian}), an infinite combination of $R$-matrices \cite{JKKKM} has to be considered. Generally speaking, infinite combinations of the $R$-matrix are not free from the difficulty of divergences. However the problem can be alternatively tackled using the $q$-vertex operators introduced by Baxter \cite{Baxter2} and Jimbo, Miwa, and Nakayashiki \cite{JMN}. Note that the CTM for $U_q(\widehat{sl_2})$ \cite{FM} gives a supporting argument for such mathematical formulation, that we recall now.

 Following the strategy of \cite{DFJMN,JKKKM}, it is possible to formulate the transfer matrix associated with (\ref{def:Hamiltonian}) using the $q$-vertex operators of $U_q(\widehat{sl_2})$.
Let $\{v_+,v_-\}$ denote the natural basis of $V={\bf C}^2$. 
Consider the infinite dimensional vector space  $\cdots \otimes V_3 \otimes V_2 \otimes V_1$   on which the Hamiltonian (\ref{def:Hamiltonian}) acts.  Let us introduce the subspace ${\cal H}^{(i)}~(i=0,1)$
of the half-infinite spin chain:
\begin{eqnarray}
{\cal H}^{(i)}=Span \{
\cdots \otimes v_{p(N)} \otimes \cdots \otimes v_{p(2)} \otimes v_{p(1)}|~p(N)=(-1)^{N+i}~(N \gg 1)\},
\end{eqnarray}
where $p:{\bf N} \to \{\pm \}$. In the thermodynamic limit, it is conjectured that level one infinite dimensional representations of $U_q(\widehat{sl_2})$ \cite{JKKKM,BB1} are embedded into ${\cal H}^{(i)}~(i=0,1)$ using 
the $q$-vertex operators. Let $V(\Lambda_i)~(i=0,1)$ denote the level one irreducible highest weight $U_q(\widehat{sl_2})$ representations with fundamental weights $\Lambda_i$. 
Let $V_\zeta$ the evaluation representation of the basic representation of $U_q({sl_2})$.
In the VOA we identify the space ${\cal H}^{(i)}$ with the space $V(\Lambda_i)$:
\begin{eqnarray}
{\cal H}^{(i)}=V(\Lambda_i).
\end{eqnarray}
Introduce the type-I vertex operators $\Phi^{(1-i,i)}_\epsilon(\zeta)$ 
as the intertwiner of $U_q(\widehat{sl_2})$ \cite{DFJMN,JKKKM}. 
\begin{eqnarray}
\Phi^{(1-i,i)}(\zeta) : V(\Lambda_i) \to V(\Lambda_{1-i})\otimes V_\zeta,
~~~
\Phi^{(1-i,i)}(\zeta)=\sum_\epsilon \Phi_\epsilon^{(1-i,i)}(\zeta)
\otimes v_\epsilon.
\end{eqnarray}
They satisfy the following commutation relation,
the duality, and the invertibility properties:
\begin{eqnarray}
&&
{\Phi}_{\epsilon_2}^{(i,1-i)}(\zeta_2)
{\Phi}_{\epsilon_1}^{(1-i,i)}(\zeta_1)
=
\sum_{\epsilon_1', \epsilon_2'=\pm}R(\zeta_1/\zeta_2)_{\epsilon_1 \epsilon_2}^{\epsilon_1' \epsilon_2'}
{\Phi}_{\epsilon_1'}^{(i,1-i)}(\zeta_1)
{\Phi}_{\epsilon_2'}^{(1-i,i)}(\zeta_2),
\label{eqn:com-math-VO}
\\
&&{\Phi}_{\epsilon}^{*(1-i,i)}(\zeta)
=
{\Phi}_{-\epsilon}^{(1-i,i)}(-q^{-1}\zeta),
\label{eqn:dual-math-VO}
\\
&&
g \sum_{\epsilon=\pm} \Phi^{* (i,1-i)}_\epsilon(\zeta)\Phi_\epsilon^{(1-i,i)}(\zeta)=id,~~~
g \Phi_{\epsilon_1}^{(i,1-i)}(\zeta)\Phi_{\epsilon_2}^{* (1-i,i)}(\zeta)=\delta_{\epsilon_1 \epsilon_2} id,
\label{eqn:inversion-math-VO}
\end{eqnarray}
where the $R-$matrix  $R(\zeta)$ is given in Appendix \ref{appendix:A}, and
\begin{eqnarray}
g=\frac{(q^2;q^4)_\infty}{(q^4;q^4)_\infty},~~~~~
(z;p)_\infty=\prod_{n=0}^\infty(1-p^nz).
\end{eqnarray}

As the generating function of the Hamiltonian $H_B^{(\pm)}$ (\ref{def:Hamiltonian}) one introduces the ``renormalized" transfer matrix ${T}_B^{(\pm,i)}(\zeta; r,s)$ using the $q$-vertex operators \cite{JKKKM,BB1}:
\begin{eqnarray}
{T}_B^{(\pm,i)}(\zeta;r,s)=g \sum_{\epsilon_1, \epsilon_2=\pm}
{\Phi}_{\epsilon_1}^{* (i,1-i)}(\zeta^{-1})K^{(\pm)}(\zeta;r,s)_{\epsilon_1}^{\epsilon_2}
{\Phi}_{\epsilon_2}^{(1-i,i)}(\zeta)
\label{def:math-transfer}
\end{eqnarray}
where the $K$-matrices $K^{(\pm)}(\zeta)$ are reported in Appendix \ref{appendix:A}. 
In the VOA we relate the Hamiltonian $H_B^{(\pm)}$ to
the ``renormalized" transfer matrix $T_B^{(\pm,i)}(\zeta;r,s)$ by
\begin{eqnarray}
\left.\frac{d}{d\zeta}T_B^{(\pm,i)}(\zeta;r,s)\right|_{\zeta=1}=
\frac{4q}{1-q^2}H_B^{(\pm)}+{\rm const.}
\end{eqnarray}
From the boundary Yang-Baxter equation satisfied by  $K^{(\pm)}(\zeta)$ and the properties of the $q$-vertex operators
(\ref{eqn:BYBE}), (\ref{eqn:BUC}), (\ref{eqn:com-math-VO}), 
(\ref{eqn:dual-math-VO}), (\ref{eqn:inversion-math-VO}),
the following properties of the ``renormalized" transfer matrix hold:
\begin{eqnarray}
&&~[{T}_B^{(\pm,i)}(\zeta_1;r,s),{T}_B^{(\pm,i)}(\zeta_2;r,s)]=0~~~~{\rm for~any}~~\zeta_1, \zeta_2,\\
&&~T_B^{(\pm,i)}(1;r,s)=id,~~~
T_B^{(\pm,i)}(\zeta;r,s)T_B^{(\pm,i)}(\zeta^{-1};r,s)=id,\\
&&~T_B^{(\pm,i)}(-q^{-1}\zeta^{-1};r,s)=T_B^{(\pm,i)}(\zeta;r,s).
\end{eqnarray}

Note that the CTM argument for $U_q(\widehat{sl_2})$ \cite{FM}, as well as the alternative support within Onsager's framework \cite{BB1},
justify the introduction of the transfer matrix (\ref{def:math-transfer}). Besides the fact that the $q$-vertex operators $\Phi_\epsilon^{(1-i,i)}(\zeta)$,  $\Phi_\epsilon^{*(1-i,i)}(\zeta)$ associated with $U_q(\widehat{sl_2})$  are well-defined objects - free from the difficulty of divergences - they are the unique solution of the intertwining relations defining the $q$-vertex operators of the $q$-Onsager algebra, which characterizes the hidden non-Abelian symmetry of (\ref{def:Hamiltonian}) (see \cite{BB1} for details).

%\vspace{3cm}

\subsection{Eigenvectors, duals and basic scalar products}
For the model (\ref{def:Hamiltonian}) with diagonal ($s=0$) or triangular ($s\neq 0$) boundary conditions, finding the eigenvalues, eigenvectors and dual eigenvectors of the Hamiltonian leads to consider the spectral problem of the transfer matrix (\ref{def:math-transfer}). In the VOA, two fundamental solutions - the so-called vacuum eigenvectors $|\pm; i\rangle_B$ ($i=0,1$) - can be constructed using bosonization. Dual vacuum eigenvectors $~_B\langle i;\pm|$   are constructed similarly. For general values $s$, by definition they satisfy
\begin{eqnarray}
T_B^{(\pm, i)}(\zeta;r,s)|\pm; i\rangle_B
&=&
\Lambda^{(i)}(\zeta;r)|\pm ;i \rangle_B,
\label{def:vacuum}\\
~_B\langle i;\pm|T_B^{(\pm, i)}(\zeta;r,s)
&=&
~_B\langle i;\pm|\Lambda^{(i)}(\zeta;r)
\label{def:dual-vacuum}\ .
\end{eqnarray}
Proposed in \cite{JKKKM} for $s=0$ and \cite{BK1} for $s\neq 0$, the structure of the eigenvectors and duals is as follows. Let $e_j,f_j,h_j$, $j=0,1$, be the Chevalley generators of $U_q(\widehat{sl_2})$. The vacuum eigenvectors read\footnote{Here we have introduced the $q$-exponential function $\exp_q(x)$ defined by $\exp_q(x)=\sum_{n=0}^\infty \frac{q^{\frac{n(n-1)}{2}}}{[n]_q!}x^n$.}:
\begin{eqnarray}
&&|+;0\rangle_B =\exp_{q^{-1}}\left(s f_0 \right)|0\rangle_B, \qquad  \qquad \quad\  |+;1\rangle_B=\exp_{q}\left(\frac{~s~}{rq}e_1 q^{-h_1}\right)|1\rangle_B,\label{vec+}\\
&& |-;0\rangle_B =\exp_{q}\left(-\frac{~s~}{q}e_0 q^{-h_0}\right)|0\rangle_B,\qquad
|-;1\rangle_B=\exp_{q^{-1}}\left(-\frac{~s~}{r}f_1 \right)|1\rangle_B,\label{vec-}
\end{eqnarray}
where $|i\rangle_B$ are the vacuum eigenvectors of the diagonal boundary $s=0$ \cite{JKKKM} (see definitions below). Dual vacuum eigenvectors take the form:
\begin{eqnarray}
&&
~_B\langle 0; +|=~_B\langle 0|\exp_q\left(-s f_0\right), \qquad\qquad \quad  \ \ ~_B\langle 1;+|=~_B\langle 1|\exp_{q^{-1}}\left(-\frac{~s~}{r q} e_1 q^{-h_1}\right),\label{dualvec+}
\\
&&~_B\langle 0; -|=~_B\langle 0|\exp_{q^{-1}}\left(\frac{~s~}{q} e_0 q^{-h_0}\right), \qquad ~_B\langle 1;-|=~_B\langle 1|\exp_{q}\left(\frac{~s~}{r} f_1\right),\label{dualvec-}
\end{eqnarray}
where $~_B\langle i|$ are the dual vacuum eigenvectors of the diagonal boundary $s=0$ \cite{JKKKM}. For $s=0$, realizations of vacuum vectors and dual ones in terms of $q$-bosons have been obtained in \cite{JKKKM}. 
See the definition of the $q$-bosons $a_n~(n \in {\bf N}_{\neq 0})$
in Appendix B.
Let us denote $|i\rangle$ (resp. $\langle i|$)  the highest (resp. dual highest) weight vector by:
\begin{eqnarray}
|i\rangle=1\otimes e^{\Lambda_i},~~~~~\langle i|=1 \otimes e^{-\Lambda_i},
\end{eqnarray}
respectively. Explicitly, for the diagonal case $s=0$ the vacuum eigenvectors and duals are given by:
\begin{eqnarray}
|i\rangle_B=\exp\left(F_i\right)|i\rangle, && F_i=\frac{1}{2}\sum_{n=1}^\infty
\frac{n \alpha_n}{[2n]_q[n]_q}a_{-n}^2+\sum_{n=1}^\infty \beta_n^{(i)}a_{-n},\label{def:vac}\\
~_B\langle i|=\langle i|\exp\left(G_i\right), && G_i=\frac{1}{2}\sum_{n=1}^\infty \frac{n \gamma_n}{[2n]_q[n]_q} a_n^2
+\sum_{n=1}^\infty \delta_n^{(i)}a_n.\label{def:vacdual}
\end{eqnarray}
Here we have set 
\begin{eqnarray}
&&
\alpha_n=-q^{6n},~~~\gamma_n=-q^{-2n},
\label{def:alpha}
\end{eqnarray}
and
\begin{eqnarray}
\beta_n^{(i)}&=&-\theta_n \frac{q^{{5n}/{2}}(1-q^n)}{[2n]_q}+\left\{
\begin{array}{cc}
-\frac{ q^{{7n}/{2}}r^n}{ [2n]_q}&~(i=0),\\
+\frac{ q^{{3n}/{2}}r^{-n}}{ [2n]_q}&~(i=1),
\end{array}
\right.\label{def:beta}
\\
\delta_n^{(i)}&=&
\theta_n \frac{q^{-{3n}/{2}}(1-q^n)}{[2n]_q}+\left\{
\begin{array}{cc}
-\frac{ q^{-{5n}/{2}}r^n}{ [2n]_q}&~(i=0),\\
+\frac{ q^{-{n}/{2}}r^{-n}}{ [2n]_q}&~(i=1),
\end{array}
\right.\label{def:delta}
\end{eqnarray}
where
\begin{eqnarray}
[n]_q=\frac{q^n-q^{-n}}{q-q^{-1}},~~~~~
\theta_n=\left\{
\begin{array}{cc}
1& {\rm for}~n~{\rm even},\\
0& {\rm for}~n~{\rm odd}.
\end{array}
\right.
\end{eqnarray}
In addition, the two fundamental eigenvalues of the transfer matrix are given by\footnote{The elliptic theta function defined by $\Theta_p(z)=(p;p)_\infty (z;p)_\infty (p/z;p)_\infty$ is introduced.}
\begin{eqnarray}
\Lambda^{(i)}(\zeta;r)=\left\{\begin{array}{cc}
1&~ (i=0)
\\
    \frac{1}{\zeta^2}\frac{\Theta_{q^4}(r\zeta^2)
\Theta_{q^4}(q^2r\zeta^{-2})}{
\Theta_{q^4}(r \zeta^{-2})
\Theta_{q^4}(q^2r \zeta^2)}, &~(i=1)
\end{array}
\right..
\end{eqnarray}

Starting from the two vacuum eigenvectors, it is possible to create
the excited states by an application of type-II vertex operators.
Recall that type-II vertex operators $\Psi^{*(1-i,i)}_\epsilon(\xi)$
are also intertwiners of $U_q(\widehat{sl_2})$.
Together with the type-I vertex operators
${\Phi}_\epsilon^{(1-i,i)}(\zeta)$, the type-II vertex operators ${\Psi}_\mu^{(i,1-i)}(\xi)$
satisfy the following commutation relation:
\begin{eqnarray}
{\Phi}_{\epsilon}^{(i,1-i)}(\zeta)
{\Psi}_{\mu}^{*(1-i,i)}(\xi)
&=&
\tau(\zeta/\xi)
{\Psi}_{\mu}^{*(i,1-i)}(\xi)
{\Phi}_{\epsilon}^{(1-i,i)}(\zeta),
\label{eqn:com-math-VO-I-II}
\end{eqnarray}
with
\begin{eqnarray}
\tau(\zeta)=\zeta^{-1}\frac{\Theta_{q^4}(q\zeta^2)}{\Theta_{q^4}(q\zeta^{-2})}.
\end{eqnarray}
This commutation relation implies that 
\begin{eqnarray}
\Psi_{\mu_1}^{* (i+N,i+N-1)}(\xi_1)\cdots \Psi_{\mu_{N-1}}^{*(i+2,i+1)}(\xi_{N-1})\Psi_{\mu_N}^{*(i+1,i)}(\xi_N)
|\pm;i\rangle_B
\end{eqnarray}
is an eigenvector of $T_B^{(\pm,i)}(\zeta;r,s)$ with eigenvalue :
\begin{eqnarray}
\Lambda^{(i)}(\zeta;r)\prod_{j=1}^N \tau(\zeta/\xi_j)\tau(\zeta \xi_j).
\end{eqnarray}
Here the number $i+N$ in the suffix $(i+N,i+N-1)$ of the $q$-vertex operator
$\Psi_{\mu_1}^{*(i+N,i+N-1)}(\xi_1)$ should be understood modulo $2$. Note that type-II vertex operators are intertwiners of the 
$q$-Onsager algebra, which also justify their use in solving (\ref{def:Hamiltonian}) \cite{BB1}.

Finally, from the relation for the $q$-exponential function $\exp_q(x)\exp_{q^{-1}}(-x)=1$ the basic scalar products between the vacuum eigenvectors and their duals can be computed. Observe:
\begin{eqnarray}
~_B\langle i; \pm|\pm; i\rangle_B=~_B\langle i|i\rangle_B.
\end{eqnarray}
Hence, according to the formulae for the norms of 
$~_B\langle i|i\rangle_B$ given in \cite{JKKKM} it follows:
\begin{eqnarray}
~_B\langle i; \pm|\pm; i\rangle_B=\left\{
\begin{array}{cc}
\frac{
(q^4r^2;q^8)_\infty}{
(q^6;q^8)_\infty (q^2r^2;q^8)_\infty}&~~(i=0),\\
\frac{
(q^4/r^2;q^8)_\infty}{
(q^6;q^8)_\infty (q^2/r^2;q^8)_\infty}&~~(i=1).
\end{array}\right.
\end{eqnarray}

\subsection{Spin-reversal property}
In the thermodynamic limit, the analogue in the VOA of the spin-reversal operator $\hat{\nu}:{\cal H}^{(i)} \longrightarrow {\cal H}^{(1-i)}~(i=0,1)$ 
is played  by the vector-space isomorphism $\nu : V(\Lambda_i)\longrightarrow V(\Lambda_{1-i})$, 
associated with the Dynkin diagram symmetry \cite{DFJMN}. 
Recall that the action of $\nu$ on the Chevalley generators of $U_q(\widehat{sl_2})$ is given by:
\begin{eqnarray}
&&
\nu e_j \nu=e_{1-j},~~
\nu f_j \nu=f_{1-j},~~
\nu q^{h_j} \nu=q^{h_{1-j}}.
\end{eqnarray}
Moreover, one has \cite{DFJMN}:
\begin{eqnarray}
\nu~\Phi_\epsilon^{(i,1-i)}(\zeta)~\nu=\Phi_{-\epsilon}^{(1-i,i)}(\zeta),\qquad  \nu~\Psi_\mu^{*(i,1-i)}(\zeta)~\nu=\Psi_{-\mu}^{*(1-i,i)}(\zeta).
\label{eqn:reversal-VO}
\end{eqnarray}
Under the action of 
$\nu$, it is possible to show that the transfer matrix (\ref{def:math-transfer}) satisfies (see \cite{JKKKM,BK1} 
for details):
\begin{eqnarray}
\nu T_B^{(\pm,1)}(\zeta;r,s) \nu=\Lambda^{(1)}(\zeta;r) T_B^{(\mp,0)}(\zeta;1/r,-s/r).
\label{eqn:reversal-transfer}
\end{eqnarray}
As a consequence, according to the structure of (\ref{vec+})-(\ref{dualvec-}) and the results for the diagonal case $s=0$ \cite{JKKKM}, 
the vacuum eigenvectors and their duals should be related by\footnote{A rigorous proof of these relations - besides the known consistency checks - in the bosonization framework remains an open problem.}:
\begin{eqnarray}
\left.\nu\left(|\pm; i \rangle_B\right)\right|_{r \to 1/r, s\to -s/r}=|\mp; 1-i\rangle_B, \label{eqn:reversal-vacuum-triangular}\\
\left.\nu\left(_B\langle i; \pm |\right)\right|_{r \to 1/r, s\to -s/r}=~_B\langle 1-i; \mp|.
\label{eqn:reversal-dual-vacuum-triangular}
\end{eqnarray}

Above relations will be used in further Sections, 
for instance to derive two integrals representations for the form factors, as well as linear the relations between multiple integrals.

\section{Form factors for a diagonal boundary}

In this Section, we study
the form factors of local operators for a {\it diagonal} boundary.
Integral representations 
of form factors of local operators are proposed using bosonization.

\subsection{Definition}

In this Section we study
the vacuum expectation values of 
the $q$-vertex operators
with natural numbers $M, N$ with $M=N~(mod.2)$ :
\begin{eqnarray}
&&F^{(i, M,N)}_{\epsilon_1,\cdots,\epsilon_M,\mu_1,\cdots,\mu_N}(\zeta_1,\cdots,\zeta_M,\xi_1,\cdots,\xi_N;r)
\nonumber\\
&=&
\frac{~_B\langle i|\Phi_{\epsilon_1}^{(i,i+1)}(\zeta_1)\cdots
\Phi_{\epsilon_{M}}^{(i+M-1,i+M)}(\zeta_{M})
\Psi_{\mu_1}^{*(i+N,i+N-1)}(\xi_1)
\cdots
\Psi_{\mu_N}^{*(i+1,i)}(\xi_N)
|i\rangle_B
}{~_B\langle i|i \rangle_B}.
\label{def:form-factor-diagonal-1}
\end{eqnarray}
Here the number $i$ in the suffix $(i,i+1)$ of
$q$-vertex operators $\Phi_\epsilon^{(i,i+1)}(\zeta)$ 
should be understood modulo 2.

Let $L$ be a linear operator on the $n$-fold tensor products
of the two-dimensional space $V_n \otimes \cdots \otimes V_2 \otimes V_1$.
The corresponding local operator ${\cal L}$ acting on $V(\Lambda_i)$
can be defined in terms of the type-I vertex operators in exactly the same way as in the bulk theory \cite{DFJMN}.
Explicitly, if $(E_{\epsilon \epsilon'})_n$ is the matrix at the $n$-th site
$E_{\epsilon \epsilon'}\otimes 
\underbrace{id \otimes \cdots \otimes id}_{n-1}$,
the corresponding local operator 
$({\cal E}_{\epsilon \epsilon'})_n$
is given by
$({\cal E}_{\epsilon \epsilon'})_n=E_{\epsilon \epsilon'}(1,1,\cdots,1)$,
where we have set
\begin{eqnarray}
E_{\epsilon \epsilon'}(\zeta_1,\zeta_2,\cdots,\zeta_n)
&=&
g^n
\sum_{\epsilon_1,\cdots,\epsilon_{n-1}=\pm}
\Phi_{\epsilon_1}^{* (i,i+1)}(\zeta_1)
\cdots 
\Phi_{\epsilon_{n-1}}^{*(i+n-2,i+n-1)}(\zeta_{n-1})
\Phi_{\epsilon}^{* (i+n-1,i+n)}(\zeta_n)\nonumber\\
&\times&
\Phi_{\epsilon'}^{(i+n,i+n-1)}(\zeta_n)
\Phi_{\epsilon_{n-1}}^{(i+n-2,i+n-1)}(\zeta_{n-1})
\cdots
\Phi_{\epsilon_1}^{(i+1,i)}(\zeta_1).
\label{def:local-operator}
\end{eqnarray}
From the inversion property of 
the $q$-vertex operators (\ref{eqn:inversion-math-VO}),
the form factors
of local operators \\
$({\cal E}_{\epsilon_m \epsilon_m'})_m
({\cal E}_{\epsilon_{m-1} \epsilon_{m-1}'})_{m-1}
\cdots
({\cal E}_{\epsilon_1 \epsilon_1'})_{1}$
with $M=2m$ and $N$ : even are given by
\begin{eqnarray}
&&
~_B\langle i|
({\cal E}_{\epsilon_m \epsilon_m'})_m
({\cal E}_{\epsilon_{m-1} \epsilon_{m-1}'})_{m-1}
\cdots
({\cal E}_{\epsilon_1 \epsilon_1'})_{1}
\Psi_{\mu_1}^{*(i+N,i+N-1)}(\xi_1)
\cdots
\Psi_{\mu_N}^{*(i+1,i)}(\xi_N)
|i\rangle_B
\nonumber\\
&&=
g^m \times ~_B\langle i|i\rangle_B \times
F_{-\epsilon_1,\cdots,-\epsilon_m,\epsilon_m',\cdots,\epsilon_1',
\mu_1,\cdots,\mu_N}
^{(i, 2m,N)}
(~\overbrace{-q^{-1},\cdots,-q^{-1}}^m, 
\overbrace{1, \cdots, 1}^m, \xi_1,\cdots, \xi_N;r).
\label{def:form-factor-diagonal-2}
\end{eqnarray}
In what follows 
we call
both (\ref{def:form-factor-diagonal-1}) and 
(\ref{def:form-factor-diagonal-2})
the form factors of local operators.
In the next Section we calculate the form factors by using bosonizations.

\subsection{Integral representation}

In this Section we focus our attention on
the half-infinite XXZ spin chain with a diagonal boundary.
We calculate the form factors with $M=N~(mod.2)$ : 
\begin{eqnarray}
&&F^{(i,M,N)}_{\epsilon_1,\cdots,\epsilon_M,\mu_1,\cdots,\mu_N}(\zeta_1,\cdots,\zeta_M,\xi_1,\cdots,\xi_N;r).
\end{eqnarray}
Here we summarize detailed calculations.
In this Section we set
\begin{eqnarray}
A=\{1\leq a \leq M|\epsilon_a=+\},~~~B=\{1\leq b \leq N|\mu_b=+\}.
\end{eqnarray}
Let $|A|$ and $|B|$ denote number of elements of the sets $A$ and $B$,
respectively.
Note that the following infinite product relation :
\begin{eqnarray}
\exp\left(-\sum_{n=1}^\infty \frac{1}{n}
\frac{z^n}{(1-p_1^n)(1-p_2^n)\cdots (1-p_N^n)}\right)
=(z;p_1,p_2,\cdots,p_N)_\infty
\end{eqnarray}
will be used, where we denote
\begin{eqnarray}
(z;p_1,p_2,\cdots,p_N)_\infty=\prod_{n_1,n_2,\cdots,n_N=0}^\infty
(1-p_1^{n_1}p_2^{n_2}\cdots p_N^{n_N}z).
\end{eqnarray}
We introduce the double-infinite products :
\begin{eqnarray}
\{z\}_\infty=(z;q^4,q^4)_\infty,~~~~~[z]_\infty=(z;q^8,q^8)_\infty.
\end{eqnarray}
Extending the analysis of \cite{BK1}, the normal ordering of products of type-I and type-II vertex operators are
given by
\begin{eqnarray}
&&
\Phi_{\epsilon_1}^{(i,i+1)}(\zeta_1)\cdots
\Phi_{\epsilon_{M}}^{(i+M-1,i+M)}(\zeta_{M})\cdot
\Psi_{\mu_1}^{*(i+N-1,i+N)}(\xi_1)\cdots 
\Psi_{\mu_N}^{*(i+1,i)}(\xi_{N})
\nonumber\\
&=&q^{|B|}(-q^3)^{\frac{1}{4}M(M-1)+\frac{1}{4}N(N-1)
-\frac{1}{2}MN-\sum_{a \in A}a-\sum_{b \in B}b}(1-q^2)^{|A|+|B|}
\prod_{j=1}^{M}
\zeta_j^{\frac{1+\epsilon_j}{2}+M-N-j}
\prod_{j=1}^{N} \xi_j^{\frac{1+\mu_j}{2}+N-j}\nonumber\\
&\times&
\prod_{1\leq j<k \leq M}\frac{(q^2\zeta_k^2/\zeta_j^2;q^4)_\infty}
{(q^4\zeta_k^2/\zeta_j^2;q^4)_\infty}
\prod_{1\leq j <k \leq N}
\frac{(\xi_k^2/\xi_j^2;q^4)_\infty}{(q^2\xi_k^2/\xi_j^2;q^4)_\infty}
\prod_{j=1}^{M}\prod_{k=1}^{N} \frac{(q^3\xi_k^2/\zeta_j^2;q^4)_\infty}{
(q\xi_k^2/\zeta_j^2;q^4)_\infty}\nonumber\\
&\times&
\oint \cdots \oint_C
\prod_{a \in A}\frac{dw_a}{2\pi \sqrt{-1}}w_a
\prod_{b \in B}\frac{du_b}{2\pi \sqrt{-1}}
\frac{
\prod_{a,b \in A
\atop{a<b}} (w_a-w_b)(w_a-q^2w_b)
\prod_{a,b \in B \atop{a<b}} (u_a-u_b)(u_a-q^{-2}u_b)
}{
\prod_{a \in A}\prod_{b \in B}(w_a-qu_b)(w_a-q^{-1}u_b)}
\nonumber\\
&\times&
\frac{
\prod_{a \in A}\prod_{j=1}^{N}(w_a-q^3\xi_j^2)
\prod_{b \in B}\prod_{j=1}^{M}(u_b-q^3 \zeta_j^2)
}{
\prod_{a \in A}
\left\{
\prod_{1\leq j \leq a}(\zeta_j^2-q^{-2}w_a)
\prod_{a \leq j \leq M}(w_a-q^4\zeta_j^2)\right\}
\prod_{b \in B}
\left\{
\prod_{1\leq j \leq b}(\xi_j^2-q^{-4}u_b)
\prod_{b \leq j \leq N}(u_b-q^2\xi_j^2)
\right\}}
\nonumber\\
&\times&
:\Phi_{-}^{(i,i+1)}(\zeta_1)\cdots \Phi_{-}^{(i+M-1,i+M)}(\zeta_{M})
\Psi_{-}^{*(i+N,i+N-1)}(\xi_1)\cdots 
\Psi_{-}^{*(i+1,i)}(\xi_{N})
\prod_{a \in A}X^-(w_a)
\prod_{b \in B}X^+(u_b):.
\end{eqnarray}
Here the integral contour $C$ 
is simple closed curve such that the $w_a~(a\in A)$ encircles
$q^4\zeta_j^2~(a\leq j \leq M)$, $qu_b, q^{-1}u_b~(b \in B)$ but not
$q^2\zeta_j^2~(1 \leq j \leq a)$,
and that the $u_b~(b \in B)$ encircles
$q^2\xi_j^2~(b \leq j \leq N)$ but not $q^4\xi_j^2~(1\leq j \leq b)$,
$qw_a, q^{-1}w_a~(a \in A)$.
Normal orderings of the basic operators are summarized in
Appendix \ref{appendix:B}.
The zero-mode $e^\alpha$ part of the operator :
$$:\Phi_{-}^{(i,i+1)}(\zeta_1)\cdots \Phi_{-}^{(i+M-1,i+M)}(\zeta_{M})
\Psi_{-}^{*(i+N,i+N-1)}(\xi_1)\cdots 
\Psi_{-}^{*(i+1,i)}(\xi_{N})
\prod_{a \in A}X^-(w_a)
\prod_{b \in B}X^+(u_b):$$
is given by 
$e^{\alpha((M-N)+2(|B|-|A|))}$.
Then, the condition for which the vacuum expectation value is non-vanishing 
reads :
\begin{eqnarray}
&&~_B\langle i|
:\prod_{j=1}^M
\Phi_{-}^{(i+j-1,i+j)}(\zeta_j)
\prod_{j=1}^N
\Psi_{-}^{*(i+j-1,i+j)}(\xi_j)
\prod_{a \in A}X^-(w_a)
\prod_{b \in B}X^+(u_b):
|i\rangle_B\neq 0,\nonumber\\
&&\Longleftrightarrow
~_B\langle i|e^{\alpha((M-N)+2(|B|-|A|))}|i\rangle_B\neq 0
\Longleftrightarrow (M-N)=2(|A|-|B|).
\end{eqnarray}
Provided the condition $(M-N)=2(|A|-|B|)$ is satisfied,
the vacuum expectation values of the $q$-vertex operators are given by
\begin{eqnarray}
&&~_B\langle i | 
\Phi_{\epsilon_1}^{(i,i+1)}(\zeta_1)\cdots
\Phi_{\epsilon_{M}}^{(i+M-1,i+M)}(\zeta_{M})
\Psi_{\mu_1}^{*(i+N,i+N-1)}(\xi_1)\cdots
\Psi_{\mu_{2N}}^{*(i+1,i)}(\xi_{N})
|i \rangle_B\nonumber\\
&=&q^{|B|}(-q^3)^{\frac{1}{4}(M-N)^2+\frac{i}{2}(M-N)
-\sum_{a \in A}a-\sum_{b \in B}b}(1-q^2)^{|A|+|B|}
\prod_{j=1}^{M}
\zeta_j^{\frac{1+\epsilon_j}{2}+M-N-j+i}
\prod_{j=1}^{N} \xi_j^{\frac{1+\mu_j}{2}+N-j+1-i}\nonumber\\
&\times&
\prod_{1\leq j<k \leq M}\frac{(q^2\zeta_k^2/\zeta_j^2;q^4)_\infty}
{(q^4\zeta_k^2/\zeta_j^2;q^4)_\infty}
\prod_{1\leq j <k \leq N}
\frac{(\xi_k^2/\xi_j^2;q^4)_\infty}{(q^2\xi_k^2/\xi_j^2;q^4)_\infty}
\prod_{j=1}^{M}\prod_{k=1}^{N} \frac{(q^3\xi_k^2/\zeta_j^2;q^4)_\infty}{
(q\xi_k^2/\zeta_j^2;q^4)_\infty}\nonumber\\
&\times&
\oint \cdots \oint_C
\prod_{a \in A}\frac{dw_a}{2\pi \sqrt{-1}}w_a^{1-i}
\prod_{a \in B}\frac{du_a}{2\pi \sqrt{-1}}u_a^i
\frac{
\prod_{a,b \in A
\atop{a<b}} (w_a-w_b)(w_a-q^2w_b)
\prod_{a,b \in B \atop{a<b}} (u_a-u_b)(u_a-q^{-2}u_b)
}{
\prod_{a \in A}\prod_{b \in B}(w_a-qu_b)(w_a-q^{-1}u_b)}\nonumber\\
&\times&
\frac{
\prod_{a \in A}\prod_{j=1}^{N}(w_a-q^3\xi_j^2)
\prod_{a \in B}\prod_{j=1}^{M}(u_a-q^3 \zeta_j^2)
}{
\prod_{a \in A}
\left\{
\prod_{1\leq j \leq a}(\zeta_j^2-q^{-2}w_a)
\prod_{a \leq j \leq M}(w_a-q^4\zeta_j^2)\right\}
\prod_{a \in B}
\left\{
\prod_{1\leq j \leq a}(\xi_j^2-q^{-4}u_a)
\prod_{a \leq j \leq N}(u_a-q^2\xi_j^2)
\right\}}
\nonumber\\
&\times&
~_B\langle i|
e^{\sum_{j=1}^{M}P(\zeta_j^2)
-\sum_{j=1}^{N}P(q^{-1}\xi_j^2)
+\sum_{a \in A}R^-(w_a)
+\sum_{a \in B}R^+(u_a)}\nonumber\\
&\times&
e^{\sum_{j=1}^{M}Q(\zeta_j^2)
-\sum_{j=1}^{N}Q(q\xi_j^2)
+\sum_{a \in A}S^-(w_a)
+\sum_{a \in B}S^+(v_a)}
|i\rangle_B.
\end{eqnarray}
Next we calculate the following vacuum expectation value explicitly.
\begin{eqnarray}
&&~_B\langle i|
e^{\sum_{j=1}^{M}P(\zeta_j^2)
-\sum_{j=1}^{N}P(q^{-1}\xi_j^2)
+\sum_{a \in A}R^-(w_a)
+\sum_{a \in B}R^+(u_a)}
\\
&\times&
e^{\sum_{j=1}^{M}Q(\zeta_j^2)
-\sum_{j=1}^{N}Q(q\xi_j^2)
+\sum_{a \in A}S^-(w_a)
+\sum_{a \in B}S^+(v_a)}
|i\rangle_B=\langle i|e^{G^{(i)}}e^{\sum_{n=1}^\infty a_{-n}X_n}
e^{-\sum_{n=1}^\infty a_n Y_n}e^{F^{(i)}}|i\rangle.\nonumber
\end{eqnarray}
Here we have set
\begin{eqnarray}
X_n&=&\frac{q^{{7 n}/{2}}}{[2 n]}
\sum_{j=1}^{M}
\zeta_j^{2n}
-\frac{q^{5n/2}}{[2n]}\sum_{j=1}^{N}\xi_j^{2n}
-\frac{q^{{n}/{2}}}{[n]}\sum_{a \in A}w_a^n+
\frac{q^{-{n}/{2}}}{[n]}\sum_{a \in B}u_a^n,
\\
Y_n&=&
\frac{q^{-{5n}/{2}}}{[2n]}\sum_{j=1}^{M}
\zeta_j^{-2n}
-\frac{q^{-7n/2}}{[2n]}
\sum_{j=1}^{N}\xi_j^{-2n}
-\frac{q^{{n}/{2}}}{[n]}
\sum_{a \in A}w_a^{-n}+\frac{q^{-{n}/{2}}}{[n]}\sum_{a \in B}
u_a^{-n}.
\end{eqnarray}
We use the following formula given in \cite{JKKKM}.
\begin{eqnarray}
&&\frac{~_B\langle i|
\exp\left(\sum_{n=1}^\infty a_{-n}X_n\right)\exp\left(-\sum_{n=1}^\infty a_n Y_n\right)|i \rangle_B}
{~_B\langle +;i|i;+ \rangle_B}
\label{eqn:vev}
\\
&=&
\exp\left(
\sum_{n=1}^\infty
\frac{[2n][n]}{n}\frac{1}{1-\alpha_n \gamma_n}
\left\{\frac{1}{2}\gamma_n X_n^2-\alpha_n \gamma_n X_n Y_n +\frac{1}{2}\alpha_n Y_n^2+(\delta_n^{(i)}
+\gamma_n \beta_n^{(i)})X_n-(\beta_n^{(i)}+\alpha_n \delta_n^{(i)})Y_n
\right\}\right).\nonumber
\end{eqnarray}
Here we have used
$\alpha_n, \gamma_n$, $\beta_n^{(i)}$, and $\delta_n^{(i)}$ in
(\ref{def:alpha}), (\ref{def:beta}), and (\ref{def:delta}), respectively.
The formulae summarized in Appendix \ref{appendix:D} are useful 
for calculations.

~\\
$\bullet$~The following is the main result of this Section. 
For $(|A|-|B|)=\frac{1}{2}(M-N)$, we have
the following integral representation of the form factors 
of local operators :
\begin{eqnarray}
&&
F^{(i,M,N)}_{\epsilon_1,\cdots,\epsilon_M,\mu_1,\cdots,\mu_N}
(\zeta_1,\cdots.\zeta_M,\xi_1,\cdots,\xi_N;r)
\nonumber\\
&=&
q^{|B|}(1-q^2)^{|B|}
(-q^3)^{\frac{1}{4}(M-N)^2+\frac{i}{2}(M-N)-\sum_{a \in A}a-\sum_{b \in B}b}
\left(\frac{\{q^6\}_\infty}{\{q^8\}_\infty}\right)^{M}
\left(\frac{\{q^4\}_\infty}{\{q^6\}_\infty}\right)^{N}
\{(q^2;q^2)_\infty\}^{|A|+|B|}\nonumber\\
&\times&
\prod_{1\leq j<k \leq M}\frac{
\{q^6\zeta_j^2 \zeta_k^2\}_\infty
\{q^2/\zeta_j^2\zeta_k^2\}_\infty
\{q^6\zeta_j^2/\zeta_k^2\}_\infty
\{q^2\zeta_k^2/\zeta_j^2\}_\infty
}{
\{q^8\zeta_j^2\zeta_k^2\}_\infty 
\{q^4/\zeta_j^2\zeta_k^2\}_\infty 
\{q^8\zeta_j^2/\zeta_k^2\}_\infty 
\{q^4\zeta_k^2/\zeta_j^2\}_\infty}
\nonumber\\
&\times&
\prod_{1\leq j<k \leq N}\frac{
\{q^4\xi_j^2 \xi_k^2\}_\infty
\{1/\xi_j^2\xi_k^2\}_\infty
\{q^4\xi_j^2/\xi_k^2\}_\infty
\{\xi_k^2/\xi_j^2\}_\infty
}{
\{q^6\xi_j^2\xi_k^2\}_\infty 
\{q^2/\xi_j^2\xi_k^2\}_\infty 
\{q^6\xi_j^2/\xi_k^2\}_\infty 
\{q^2\xi_k^2/\xi_j^2\}_\infty}\nonumber\\
&\times&
\prod_{j=1}^{M}
\prod_{k=1}^{N}
\frac{
\{q^7\zeta_j^2 \xi_k^2\}_\infty
\{q^3/\zeta_j^2 \xi_k^2\}_\infty
\{q^7 \zeta_j^2/\xi_k^2\}_\infty
\{q^3 \xi_k^2/\zeta_j^2\}_\infty
}{
\{q^5 \zeta_j^2\xi_k^2\}_\infty 
\{q/ \zeta_j^2\xi_k^2\}_\infty 
\{q^5 \zeta_j^2/\xi_k^2\}_\infty 
\{q \xi_k^2/\zeta_j^2\}_\infty}\nonumber\\
&\times&
\prod_{j=1}^{M} 
\frac{[q^{10}\zeta_j^4]_\infty [q^{14}\zeta_j^4]_\infty 
[q^{10}/\zeta_j^4]_\infty [q^6 /\zeta_j^4]_\infty}{
[q^{12}\zeta_j^4]_\infty [q^{16}\zeta_j^4]_\infty 
[q^{12}/\zeta_j^4]_\infty [q^8/\zeta_j^4]_\infty}
\prod_{j=1}^{N} 
\frac{[q^{4}\xi_j^4]_\infty [q^{8}\xi_j^4]_\infty 
[1/\xi_j^4]_\infty [q^4 /\xi_j^4]_\infty}{
[q^{6}\xi_j^4]_\infty [q^{10}\xi_j^4]_\infty 
[q^{2}/\xi_j^4]_\infty [q^{6}/\xi_j^4]_\infty}
\nonumber
\\
&\times&
\prod_{j=1}^{M}\zeta_j^{\frac{1+\epsilon_j}{2}+M-N-j+i}
\prod_{j=1}^{N}\xi_j^{\frac{\mu_j+1}{2}+N-j+1-i}
\oint \cdots \oint_{C^{(i)}}
\prod_{a \in A}\frac{dw_a}{2\pi \sqrt{-1}}w_a^{1-i}
\prod_{b \in B}\frac{du_b}{2\pi \sqrt{-1}}u_b^i
\nonumber\\
%%%%%%%%%%%%%%%%%%%%%%%%%%%%%%%%%%%%%%%%%%%%%%%%%%%%%%%%%%
&\times&
\frac{
\prod_{a \in A}(w_a^2/q^2;q^4)_\infty (q^6/w_a^2;q^4)_\infty
\prod_{b \in B}(u_b^2/q^2;q^4)_\infty (q^6/u_b^2;q^4)_\infty}
{
\prod_{a \in A} \prod_{b \in B}
w_a^2 (w_au_b/q^3;q^2)_\infty 
(q^3 w_a/u_b;q^2)_\infty 
(u_b/qw_a;q^2)_\infty 
(q^5/w_a u_b;q^2)_\infty}
\nonumber\\
%%%%%%%%%%%%%%%%%%%%%%%%%%%%%%%%%%%%%%%%%%%%%%%%%%%%%%%%%%
&\times&
\frac{ \prod_{a,b \in A
\atop{a<b}}w_a^2
(w_aw_b/q^2;q^2)_\infty 
(q^4w_a/w_b;q^2)_\infty 
(w_b/w_a;q^2)_\infty 
(q^6/w_aw_b;q^2)_\infty}
{
\prod_{a \in A}
\left\{\prod_{1\leq j \leq a}(\zeta_j^2-q^{-2}w_a)
\prod_{a\leq j \leq M}(w_a-q^4 \zeta_j^2)\right\}
}\nonumber\\
&\times&
\frac{
\prod_{a,b \in B
\atop{a<b}}u_a^2
(u_au_b/q^4;q^2)_\infty 
(q^2u_a/u_b;q^2)_\infty 
(u_b/q^2u_a;q^2)_\infty 
(q^4/u_au_b;q^2)_\infty}{
\prod_{b \in B}
\left\{\prod_{1\leq j \leq b}(\xi_j^2-q^{-4}u_b)
\prod_{b \leq j \leq M}(u_b-q^2 \xi_j^2)\right\}}
\nonumber\\
&\times&
\frac{
\prod_{j=1}^{M}
\prod_{b \in B} 
u_b(q\zeta_j^2 u_b;q^4)_\infty (q^3\zeta_j^2/u_b;q^4)_\infty 
(qu_b/\zeta_j^2;q^4)_\infty (q^3/\zeta_j^2 u_b;q^4)_\infty
}{
\prod_{j=1}^{M}
\prod_{a \in A}
(q^2 \zeta_j^2 w_a;q^4)_\infty 
(q^8 \zeta_j^2/w_a;q^4)_\infty (q^2w_a/\zeta_j^2;q^4)_\infty 
(q^4/\zeta_j^2 w_a;q^4)_\infty}
\nonumber
\\
&\times&
\frac{
\prod_{j=1}^{N}
\prod_{a \in A} 
w_a (q\xi_j^2 w_a;q^4)_\infty 
(q^3\xi_j^2/w_a;q^4)_\infty 
(q w_a/\xi_j^2;q^4)_\infty 
(q^3/\xi_j^2 w_a;q^4)_\infty
}{
\prod_{j=1}^{N}
\prod_{b \in B}
(\xi_j^2 u_b;q^4)_\infty 
(q^6 \xi_j^2/u_b;q^4)_\infty 
(u_b/\xi_j^2;q^4)_\infty 
(q^2/\xi_j^2 u_b;q^4)_\infty}
\nonumber\\
&\times&
\left\{\begin{array}{cc}
\prod_{j=1}^{M}\frac{(q^2r\zeta_j^2;q^4)_\infty}{(q^4r\zeta_j^2;q^4)_\infty}
\prod_{j=1}^{N}\frac{(q^3r \xi_j^2;q^4)_\infty}{
(qr\xi_j^2;q^4)_\infty}
\frac{ \prod_{b \in B}(1-ru_b/q^3)}{
 \prod_{a \in A}(1-rw_a/q^2)}& (i=0),\\
\prod_{j=1}^{M}\frac{(1/r\zeta_j^2;q^4)_\infty}{(q^2/r\zeta_j^2;q^4)_\infty}
\prod_{j=1}^{N}
\frac{(q/r\xi_j^2;q^4)_\infty}{(1/qr\xi_j^2;q^4)_\infty}
\frac{ 
\prod_{b \in B}(1-q/ru_b)}{
\prod_{a \in A}(1-q^2/rw_a)}& (i=1).
\end{array}\right.
\end{eqnarray}
Here the integration contour $C^{(0)}$ is a closed curve that satisfies 
the following conditions for $s=0,1,2,\cdots$.
The $w_a~(a\in A)$ encircles
$q^{8+4s}\zeta_j^2~(1\leq j < a)$,
$q^{4+4s}\zeta_j^2~(a \leq j \leq M)$,
$q^{4+4s}/\zeta_j^2~(1\leq j \leq M)$
$q^{-1+2s}u_b,~ q^{5+2s}/u_b~(b \in B)$,
but not
$q^{2-4s}\zeta_j^2~(1\leq j \leq a)$,
$q^{-2-4s}\zeta_j^2~(a <j \leq M)$,
$q^{-2-4s}/\zeta_j^2~(1\leq j \leq M)$,
$q^{-3-2s}u_b,~ q^{3-2s}/u_b~(b \in B)$, and $q^2/r$.
The $u_b~(b \in B)$ encircles
$q^{6+4s}\xi_j^2~(1\leq j <b)$,
$q^{2+4s}\xi_j^2~(b \leq j \leq N)$,
$q^{2+4s}/\xi_j^2~(1\leq j \leq N)$,
$q^{3+2s}w_a,~ q^{5+2s}/w_a~(a \in B)$,
but not
$q^{4-4s}\xi_j^2~(1\leq j \leq b)$,
$q^{-4s}\xi_j^2~(b < j \leq N)$,
$q^{-4s}/\xi_j^2~(1\leq j \leq N)$,
$q^{1-2s}w_a,~ q^{-3-2s}/w_a~(a \in A)$.
Here the integration contour $C^{(1)}$ is a closed curve such that $w_a~(a \in A)$ encircles
$q^2/r$ in addition the same points $C^{(0)}$ does.

\subsection{Identities between multiple integrals}

From the spin-reversal properties in (\ref{eqn:reversal-VO}), 
(\ref{eqn:reversal-vacuum-triangular}), and 
(\ref{eqn:reversal-dual-vacuum-triangular}),
the following relation between form factors of local operators holds:
\begin{eqnarray}
F^{(i,M,N)}_{\epsilon_1,\cdots,\epsilon_M,\mu_1,\cdots,\mu_N}
(\zeta_1,\cdots,\zeta_M,\xi_1,\cdots,\xi_N;r)
=
F^{(1-i,M,N)}_{-\epsilon_1,\cdots,-\epsilon_M,-\mu_1,\cdots,-\mu_N}
(\zeta_1,\cdots,\zeta_M,\xi_1,\cdots,\xi_N;1/r).
\label{eqn:reversal-diagonal-form}
\end{eqnarray}
We should understand this spin-reversal property
(\ref{eqn:reversal-diagonal-form}) as an analytic continuation of 
the parameter $r$.
From the above (\ref{eqn:reversal-diagonal-form}),
we obtain infinitely many relations between $n$-fold integrals
which cannot be reduced to the relations
between $n$-fold integrals of elliptic gamma functions 
summarized in \cite{Rains}.
Here we focus on two simple examples.\\
$\bullet$~First, from
\begin{eqnarray}
F^{(0,M,N)}_{+,\cdots,+,+,\cdots,+}
(\zeta_1,\cdots,\zeta_M,\xi_1,\cdots,\xi_N;r)=
F^{(1,M,N)}_{-,\cdots,-,-,\cdots,-}
(\zeta_1,\cdots,\zeta_M,\xi_1,\cdots,\xi_N;1/r),
\end{eqnarray}
we obtain an identity for multiple integrals of infinite products.
Here we consider the case $M=N$:
\begin{eqnarray}
&&
\prod_{j=1}^M
\frac{
(q^4r\zeta_j^2;q^4)_\infty
(r/\zeta_j^2;q^4)_\infty 
(qr\xi_j^2;q^4)_\infty 
(qr/\xi_j^2;q^4)_\infty}{
(q^2r\zeta_j^2;q^4)_\infty 
(q^2r/\zeta_j^2;q^4)_\infty
(q^3r\xi_j^2;q^4)_\infty 
(r/q\xi_j^2;q^4)_\infty}\nonumber\\
&=&
q^M (1-q^2)^M
q^{-3M(M+1)} (q^2;q^2)_\infty^{2M}\oint \cdots \oint_{C^{(0)}}
\prod_{a=1}^M \frac{dw_a}{2\pi i}w_a \prod_{b=1}^M \frac{du_b}{2\pi i}
\prod_{a=1}^M \frac{(1-ru_a/q^3)}{(1-rw_a/q^2)}\nonumber\\
&\times&
\frac{
\prod_{a=1}^M
(w_a^2/q^2;q^4)_\infty
(q^6/w_a^2;q^4)_\infty
(u_a^2/q^2;q^4)_\infty
(q^6/u_a^2;q^4)_\infty}{
\prod_{a=1}^M \prod_{b=1}^M
w_a^2 
(w_au_b/q^3;q^2)_\infty 
(q^3w_a/u_b;q^2)_\infty
(u_b/qw_a;q^2)_\infty 
(q^5/w_au_b;q^2)_\infty
}\nonumber\\
&\times&
\frac{
\prod_{1\leq a<b \leq M}
w_a^2 
(w_aw_b/q^2;q^2)_\infty 
(q^4w_a/w_b;q^2)_\infty 
(w_b/w_a;q^2)_\infty 
(q^6/w_aw_b;q^2)_\infty
}{
\prod_{a=1}^M
\left\{\prod_{1\leq j \leq a}(\zeta_j^2-q^{-2}w_a)
\prod_{a \leq j \leq M}(w_a-q^4\zeta_j^2)\right\}
}
\nonumber
\\
&\times&
\frac{
\prod_{1\leq a<b \leq M}
u_a^2 
(u_au_b/q^4;q^2)_\infty 
(q^2u_a/u_b;q^2)_\infty 
(u_b/q^2u_a;q^2)_\infty 
(q^4/u_au_b;q^2)_\infty
}{
\prod_{a=1}^M
\left\{\prod_{1\leq j \leq a}(\xi_j^2-q^{-4}u_a)
\prod_{a \leq j \leq M}(u_a-q^2\xi_j^2)\right\}}
\nonumber\\
&\times&
\frac{
\prod_{j=1}^M\prod_{b=1}^M u_b 
(q\zeta_j^2u_b;q^4)_\infty 
(q^3\zeta_j^2/u_b;q^4)_\infty
(qu_b/\zeta_j^2;q^4)_\infty
(q^3/\zeta_j^2u_b;q^4)_\infty
}{
\prod_{j=1}^M \prod_{a=1}^M
(q^2\zeta_j^2w_a;q^4)_\infty 
(q^8\zeta_j^2/w_a;q^4)_\infty 
(q^2w_a/\zeta_j^2;q^4)\infty 
(q^4/\zeta_j^2w_a;q^4)_\infty}\nonumber\\
&\times&
\frac{
\prod_{j=1}^M\prod_{a=1}^M w_a 
(q\xi_j^2w_a;q^4)_\infty 
(q^3\xi_j^2/w_q;q^4)_\infty
(qw_a/\xi_j^2;q^4)_\infty
(q^3/\xi_j^2w_a;q^4)_\infty
}{
\prod_{j=1}^M \prod_{b=1}^M
(\xi_j^2u_b;q^4)_\infty 
(q^6\xi_j^2/u_b;q^4)_\infty 
(u_b/\xi_j^2;q^4)\infty 
(q^2/\xi_j^2u_b;q^4)_\infty}.
\end{eqnarray}
The number of integrals in the RHS is $(M+N)$ where the LHS is an infinite product 
without integral.
Here the integration contour $C^{(0)}$ is given in the previous Section.
\\
$\bullet$~
Secondly, from
\begin{eqnarray}
F^{(1,2m,0)}_{\underbrace{-,\cdots,-}_{m},\underbrace{+,\cdots,+}_{m}}
(\zeta_1,\cdots,\zeta_{2m};r)=
F^{(0,2m,0)}_{\underbrace{+,\cdots,+}_{m},\underbrace{-,\cdots,-}_{m}}
(\zeta_1,\cdots,\zeta_{2m};1/r),
\end{eqnarray}
we obtain the following identity between multiple integrals of infinite products.
The numbers of integrals on both sides are $m$.
\begin{eqnarray}
&&
\oint \cdots \oint_{C^{(0)}}
\prod_{a=1}^m\frac{dw_a}{2\pi \sqrt{-1}}w_a
\prod_{j=1}^{2m}
\frac{(q^2\zeta_j^2/r;q^4)_\infty}{
(q^4\zeta_j^2/r;q^4)_\infty}
\prod_{a=1}^m \frac{1}{(1-w_a/q^2r)}\nonumber
\\
&&\times
\frac{
\prod_{a=1}^m (w_a^2/q^2;q^4)_\infty 
(q^6/w_a^2;q^4)_\infty}{
\prod_{j=1}^{2m}\prod_{a=1}^m 
(q^2\zeta_j^2w_a;q^4)_\infty 
(q^8\zeta_j^2/w_a;q^4)_\infty 
(q^2w_a/\zeta\j^2;q^4)_\infty 
(q^4/\zeta_jw_a;q^4)_\infty}\nonumber
\\
&&\times
\frac{
\prod_{1\leq a<b \leq m}w_a^2
(w_aw_b/q^2;q^2)_\infty 
(q^4w_a/w_b;q^2)_\infty
(w_b/w_a;q^2)_\infty 
(q^6/w_aw_b;q^2)_\infty
}{
\prod_{a=1}^m
\prod_{1\leq j \leq a}(\zeta_j^2-q^{-2}w_a)
\prod_{a \leq j \leq 2m}(w_a-q^4\zeta_j^2)}
\nonumber
\\
&=&(-q^3)^{-m^2+m}\prod_{j=m+1}^{2m}\zeta_j^2
\oint \cdots \oint_{C^{(1)}}
\prod_{a=m+1}^{2m}
\frac{dw_a}{2\pi \sqrt{-1}}
\prod_{j=1}^{2m}
\frac{(1/r\zeta_j^2;q^4)_\infty}{
(q^2/r\zeta_j^2;q^4)_\infty}
\prod_{a=m+1}^{2m} \frac{1}{(1-q^2/rw_a)}\nonumber
\\
&&\times
\frac{
\prod_{a=m+1}^{2m} (w_a^2/q^2;q^4)_\infty 
(q^6/w_a^2;q^4)_\infty}{
\prod_{j=1}^{2m}\prod_{a=m+1}^{2m} 
(q^2\zeta_j^2w_a;q^4)_\infty 
(q^8\zeta_j^2/w_a;q^4)_\infty 
(q^2w_a/\zeta_j^2;q^4)_\infty 
(q^4/\zeta_jw_a;q^4)_\infty}\nonumber
\\
&&\times
\frac{
\prod_{m+1\leq a<b \leq 2m}w_a^2
(w_aw_b/q^2;q^2)_\infty 
(q^4w_a/w_b;q^2)_\infty
(w_b/w_a;q^2)_\infty 
(q^6/w_aw_b;q^2)_\infty
}{
\prod_{a=m+1}^{2m}
\prod_{1\leq j \leq a}(\zeta_j^2-q^{-2}w_a)
\prod_{a \leq j \leq 2m}(w_a-q^4\zeta_j^2)}.
\end{eqnarray}
Here the integration contour $C^{(0)}$ is given in the previous Section.
In Appendix C,
we give a direct proof for the simplest case where
$\zeta_1=-q^{-1}\zeta$, $\zeta_2=\zeta$, and $m=1$
as a supporting argument :
\begin{eqnarray}
F_{-,+}^{(1,2,0)}(-q^{-1}\zeta,\zeta;r)=F_{+,-}^{(0,2,0)}(-q^{-1}\zeta,\zeta;1/r).
\end{eqnarray}

\subsection{Comparison with QISM}

In this Section, upon specializations
the previous expressions are compared with the known results obtained within the 
QISM \cite{KZMNST}. In particular, let us consider the following integral
representations of correlation functions :
\begin{eqnarray}
&&
g^m F^{(i,2m,0)}
_{-\epsilon_1,\cdots,-\epsilon_m,\epsilon_m',\cdots,\epsilon_1'}
(\overbrace{-q^{-1}\zeta,\cdots, -q^{-1}\zeta}^{m},
\overbrace{\zeta,\cdots,\zeta}^{m};r)\nonumber\\
&=&(-1)^{-\frac{1}{2}m^2-\frac{1}{2}m-\sum_{a \in A}a+(m+1)|B|}
q^{\frac{1}{2}m^2+\frac{1}{2}m
-2m|B|-\sum_{a \in A}a+\sum_{b \in B}b}
\zeta^{2(m^2-m|B|-\sum_{a \in A}a)}
\nonumber\\
&\times&
\frac{(q^2;q^2)_\infty^{2m^2-m}}{(-q^2;q^2)_\infty^m}\left\{
\frac{\Theta_{q^4}(\zeta^4)}{1-\zeta^4}\right\}^m
\{(q^2\zeta^4;q^2)_\infty
(q^2/\zeta^4;q^2)_\infty\}^{\frac{m(m-1)}{2}}(1-r\zeta^2)^m
\oint \cdots \oint_{\widetilde{C}^{(i)}}
\prod_{a \in A}\frac{dw_a}{2\pi \sqrt{-1}}
\prod_{b \in B}\frac{du_b}{2\pi \sqrt{-1}}\nonumber\\
&\times&\prod_{a \in A}
\frac{w_a^{-2m+a}\Theta_{q^2}(w_a/q)\Theta_{q^2}(-w_a/q)
(1-q^2/\zeta^2w_a)^m (1-w_a/\zeta^2)^{m-a} (1-q^2\zeta^2/w_a)^{a-1}}{
\{\Theta_{q^2}(\zeta^2w_a)\Theta_{q^2}(w_a/\zeta^2)\}^m (1-rw_a/q^2)}\nonumber\\
&\times&
\prod_{b \in B}
\frac{u_b^{-m}\Theta_{q^2}(u_b/q)\Theta_{q^2}(-u_b/q)
(1-q^2/\zeta^2 u_b)^m (1-q^2\zeta^2 /u_b)^{b-1} (1-q^4\zeta^2/u_b)^{m-b}}{
\{\Theta_{q^2}(\zeta^2 u_b)\Theta_{q^2}(u_b/\zeta^2)\}^m (1-ru_b/q^2)}\nonumber\\
&\times&
\prod_{a<b \atop{a,b \in A}}
\frac{w_a^2 w_b^2 \Theta_{q^2}(w_aw_b)\Theta_{q^2}(w_a/w_b)}{(1-q^2w_a/w_b)(1-q^4/w_aw_b)}
\prod_{a>b \atop{a,b \in B}}
\frac{u_a^2 u_b^2 \Theta_{q^2}(u_au_b)\Theta_{q^2}(u_a/u_b)}{(1-q^2u_a/u_b)(1-q^4/u_au_b)}\nonumber\\
&\times&
\prod_{a \in A}\prod_{b \in B}
\frac{w_a^2 u_b^2 \Theta_{q^2}(w_au_b)\Theta_{q^2}(w_a/u_b)}{(1-q^2w_a/u_b)(1-q^4/w_au_b)}.
\end{eqnarray}
In this Section we set
$A=\{1\leq j \leq n|\epsilon_j=-\}$, $B=\{1\leq j \leq |\epsilon_j'=+\}$.
Here the integration contour $\widetilde{C}^{(0)}$ is a closed curve that satisfies
the following conditions for $s=0,1,2,\cdots$.
The $w_a (a \in A)$ encircles
$q^{2+2s}\zeta^2$, $q^{4+2s}/\zeta^2$, but not
$q^{-2-2s}\zeta^2$, $q^{-2s}/\zeta^2$, and $q^2/r$.
The $u_b (b \in B)$ encircles
$q^{4+2s}\zeta^2$, $q^{4+2s}/\zeta^2$, but not
$q^{-2s}\zeta^2$, $q^{-2s}/\zeta^2$, and $q^2/r$.
The integral contour $\widetilde{C}^{(1)}$ is a closed curve such that $w_a (a \in A)$ and $u_b (b \in B)$
encircle $q^2/r$ in addition the same points $\widetilde{C}^{(0)}$ does.

In what follows we study integral representations upon the specialization of the
spectral parameter $\zeta=1$.
We change the variables as follows.
\begin{eqnarray}
q=e^{-\eta}~(\eta>0),~~~w_a/q=e^{2\sqrt{-1}\alpha_a},~~~u_b/q=e^{2\sqrt{-1}\delta_b},~~~r=e^{2\xi}~(\xi \in {\bf R}).
\end{eqnarray}
We use the following notation of the elliptic theta functions in \cite{WW}.
\begin{eqnarray}
\vartheta_1(u,q)&=&2 q^{\frac{1}{4}} (q^2;q^2)_\infty \sin u 
\prod_{n=1}^\infty (1-2q^{2n}\cos 2u+q^{4n}),\\
\vartheta_2(u,q)&=& 2q^{\frac{1}{4}} (q^2;q^2)_\infty \cos u
\prod_{n=1}^\infty (1+2q^{2n}\cos 2u+q^{4n}),\\
\vartheta_3(u,q)&=& (q^2;q^2)_\infty \prod_{n=1}^\infty (1+2q^{2n-1}\cos 2u+q^{4n-2}),\\
\vartheta_4(u,q)&=& (q^2;q^2)_\infty \prod_{n=1}^\infty (1-2q^{2n-1}\cos 2u+q^{4n-2}).
\end{eqnarray}
The theta function $\Theta_p(w)=(p;p)_\infty (w;p)_\infty (p/w;p)_\infty$ 
is related to $\vartheta_j(\alpha,q)$ as follows. 
\begin{eqnarray}
\Theta_{q^2}(w)=\vartheta_4(\alpha,q),~~~
\Theta_{q^2}(w/q)=
\frac{\sqrt{-1} q^{\frac{1}{4}}}{\sqrt{w}}\vartheta_1
(\alpha,q),~~~
\Theta_{q^2}(-w/q)=
\frac{q^{\frac{1}{4}}}{\sqrt{w}}\vartheta_2(\alpha,q).
\end{eqnarray}
Here we have set
$w/q=e^{2\sqrt{-1}\alpha}$.\\
\\
$\bullet$~
The following is the main result of this Section.
For $|A|+|B|=m$, we have the following integral representation
of correlation functions :
\begin{eqnarray}
&&
\frac{~_B\langle i |(E_{\epsilon_1,\epsilon_1'})_1
(E_{\epsilon_2,\epsilon_2'})_2 \cdots 
(E_{\epsilon_m,\epsilon_m'})_m|i\rangle_B}
{~_B\langle i|i\rangle_B}\nonumber\\
&=&
g^m F^{(i,2m,0)}_{-\epsilon_1,\cdots,-\epsilon_m,\epsilon_m',\cdots,\epsilon_1'}(\underbrace{-q^{-1},\cdots,-q^{-1}}_m,
\underbrace{1,\cdots,1}_m;r)\nonumber\\
&=&
2^{m(m+1)}
\sinh^m(\xi)\times
(q^2;q^2)_\infty^{m(3m+1)}(-q^2;q^2)_\infty^{2m} \nonumber\\
&\times&
(-1)^{\frac{m(m-1)}{2}+m(|B|-|A|)}
q^{-\frac{m(m-1)}{4}+m(m-\frac{1}{2})}
\int \cdots \int_{\widetilde{C}^{(i)}}
\prod_{a \in A}
\frac{d\alpha_a}{2\pi}\sqrt{-1}
\prod_{b \in B}
\frac{d\delta_b}{2\pi \sqrt{-1}}
\nonumber\\
&\times&
\prod_{a \in A}\frac{
\vartheta_1(\alpha_a,q)\vartheta_2(\alpha_a,q)~
\sin^{m+a-1}\left(\alpha_a+\frac{\sqrt{-1}}{2}\eta\right)
\sin^{m-a}(\alpha_a)}{\sin\left(\alpha_a+\frac{\sqrt{-1}}{2}\eta+\sqrt{-1}\xi
\right)~\vartheta_4(\alpha_a,q)^{2n}}
\nonumber\\
&\times&
\prod_{b \in B}
\frac{\vartheta_1(\delta_b,q)\vartheta_2(\delta_b,q)~
\sin^{m+b-1}\left(\delta_b+\frac{\sqrt{-1}}{2}\eta\right)
\sin^{m-b}\left(\delta_b+\frac{3\sqrt{-1}}{2}\eta\right)}
{\sin\left(\delta_b+\frac{\sqrt{-1}}{2}\eta+\sqrt{-1}\xi
\right)~\vartheta_4(\delta_b,q)^{2m}}\nonumber\\
&\times&
\prod_{a<b \atop{a,b \in A}}
\frac{\vartheta_1(\alpha_a+\alpha_b,q)
\vartheta_1(\alpha_a-\alpha_b,q)}{
\sin(\alpha_a+\alpha_b+\sqrt{-1}\eta)
\sin(\alpha_a-\alpha_b-\sqrt{-1}\eta)}
\prod_{a>b \atop{a,b \in B}}
\frac{\vartheta_1(\delta_a+\delta_b,q)
\vartheta_1(\delta_a-\delta_b,q)}{
\sin(\delta_a+\delta_b+\sqrt{-1}\eta)
\sin(\delta_a-\delta_b-\sqrt{-1}\eta)}
\nonumber\\
&\times&
\prod_{a \in A}
\prod_{b \in B}
\frac{\vartheta_1(\alpha_a+\delta_b,q)
\vartheta_1(\alpha_a-\delta_b,q)}{
\sin(\alpha_a+\delta_b+\sqrt{-1}\eta)
\sin(\alpha_a-\delta_b-\sqrt{-1}\eta)}.
\label{eqn:correlator}
\end{eqnarray}
Note that we obtain the same formula as the one of Lyon group \cite{KZMNST} up to a constant
$(-1)^{m(|B|-|A|)}
q^{m(m-\frac{1}{2})}$.
\begin{eqnarray}
``(\ref{eqn:correlator}) {\rm ~of~this~paper}"
=(-1)^{m(|B|-|A|)}
q^{m(m-\frac{1}{2})}
\times ``(6.18) {\rm ~of~\cite{KZMNST}}".
\end{eqnarray}

\section{Form factors for a triangular boundary}

In this Section,
we study
the form factors of local operators for a {\it triangular} boundary.
Two integral representations 
of form factors of local operators are proposed using bosonization.

\subsection{Definition}

In this Section we study
the vacuum expectation values of 
the $q$-vertex operators
with natural numbers $M, N$ with $M=N~(mod.2)$ :
\begin{eqnarray}
&&F^{(i,\pm,M,N)}_{\epsilon_1,\cdots,\epsilon_M,\mu_1,\cdots,\mu_N}(\zeta_1,\cdots,\zeta_M,\xi_1,\cdots,\xi_N;r,s)
\nonumber\\
&=&
\frac{~_B\langle i;\pm|\Phi_{\epsilon_1}^{(i,i+1)}(\zeta_1)\cdots
\Phi_{\epsilon_{M}}^{(i+M-1,i+M)}(\zeta_{M})
\Psi_{\mu_1}^{*(i+N,i+N-1)}(\xi_1)
\cdots
\Psi_{\mu_N}^{*(i+1,i)}(\xi_N)
|\pm;i\rangle_B
}{~_B\langle i;\pm|\pm;i \rangle_B}.
\label{def:form-factor-triangular-1}
\end{eqnarray}
Here the number $i$ in the suffix $(i,i+1)$ of
$q$-vertex operators $\Phi_\epsilon^{(i,i+1)}(\zeta)$ 
should be understood modulo 2.
Let us consider the form factors
of local operators
with $M=2m$ and $N$ : even given by
\begin{eqnarray}
&&
~_B\langle i;\pm|
({\cal E}_{\epsilon_m \epsilon_m'})_m
({\cal E}_{\epsilon_{m-1} \epsilon_{m-1}'})_{m-1}
\cdots
({\cal E}_{\epsilon_1 \epsilon_1'})_{1}
\Psi_{\mu_1}^{*(i+N,i+N-1)}(\xi_1)
\cdots
\Psi_{\mu_N}^{*(i+1,i)}(\xi_N)
|\pm;i\rangle_B
\label{def:form-factor-triangular-2}
\\
&&=
g^m \times ~_B\langle i;\pm|\pm;i\rangle_B \times
F_{-\epsilon_1,\cdots,-\epsilon_m,\epsilon_m',\cdots,\epsilon_1',
\mu_1,\cdots,\mu_N}
^{(\pm,i,M,N)}
(~\underbrace{-q^{-1},\cdots,-q^{-1}}_m, 
\underbrace{1, \cdots, 1}_m, \xi_1,\cdots, \xi_N;r,s).
\nonumber
\end{eqnarray}
In what follows 
we call
both (\ref{def:form-factor-triangular-1}) and 
(\ref{def:form-factor-triangular-2})
the form factors of local operators.
The form factors are calculated using bosonizations.

\subsection{First integral representation}

In this Section we focus our attention on
the half-infinite XXZ spin chain with a upper-triangular boundary.
We calculate the form factors with $M=N~(mod.2)$ : 
\begin{eqnarray}
&&F^{(+,i,M,N)}_{\epsilon_1,\cdots,\epsilon_M,\mu_1,\cdots,\mu_N}(\zeta_1,\cdots,\zeta_M,\xi_1,\cdots,\xi_N;r,s).
\end{eqnarray}
Calculations in this Section are similar to those for diagonal boundary
conditions.
In what follows we set
$A=\{1\leq a \leq M|\epsilon_a=+\}$, $B=\{1\leq b \leq N|\mu_b=+\}$,
and 
\begin{eqnarray}
L=|A|-|B|+\frac{1}{2}(N-M).
\end{eqnarray}
In the following, we understand $L=0$ as a sufficient condition
such that the expressions for triangular boundary conditions coincide with 
those for diagonal boundary conditions. 

The following bosonizations of the Chevalley generators are used:
\begin{eqnarray}
f_0=q^{\partial} \oint \frac{dv}{2\pi\sqrt{-1}}X^+(v),~~~
e_1q^{-h_1}=\oint \frac{dv}{2\pi \sqrt{-1}}X^+(v)q^{-\partial}.
\end{eqnarray}
The bosonic parts of the following operators :
\begin{eqnarray}
&&
\exp_q(-s f_0)\cdot \Phi_{\epsilon_1}^{(0,1)}(\zeta_1)\cdots
\Phi_{\epsilon_{M}}^{(M-1,M)}(\zeta_{M})\cdot
\Psi_{\mu_1}^{*(N,N-1)}(\xi_1)\cdots 
\Psi_{\mu_{N}}^{*(1,0)}(\xi_{N})
\cdot \exp_{q^{-1}}(s f_0),\\
&&
\exp_{q^{-1}}\left(-\frac{s}{r q} e_1 q^{-h_1}\right)
\cdot \Phi_{\epsilon_1}^{(1,0)}(\zeta_1)\cdots
\Phi_{\epsilon_{M}}^{(M,M+1)}(\zeta_{M})\cdot
\Psi_{\mu_1}^{*(N+1,N)}(\xi_1)\cdots 
\Psi_{\mu_N}^{*(0,1)}(\xi_{N})\cdot
\exp_{q}\left(\frac{s}{rq} e_1q^{-h_1}\right),\nonumber
\end{eqnarray}
are given by
\begin{eqnarray}
:\prod_{j=1}^M \Phi_{-}^{(i+j-1,i+j)}(\zeta_j)
\prod_{j=1}^N \Psi_-^{*(i+N-j+1,i+N-j)}(\xi_j)
\prod_{a \in A}X^-(w_a)
\prod_{a=1}^n X^+(v_a)
\prod_{a \in B}X^+(u_a):
q^{(1-2i)n \partial},
\end{eqnarray}
for $n=0,1,2,\cdots$.
Here the parameter $n$ comes from the expansion of the $q$-exponential.
The zero-mode of the above operator is
$e^{\alpha (n-|A|+|B|+\frac{1}{2}(M-N))}$,
hence the vacuum expectation value becomes non-zero i.e.:
\begin{eqnarray}
~_B\langle +;i|:\prod_{j=1}^M \Phi_{-}^{(i+j-1,i+j)}(\zeta_j)
\prod_{j=1}^N \Psi_-^{*(i+N-j+1,i+N-j)}(\xi_j)
\prod_{a \in A}X^-(w_a)
\prod_{a=1}^n X^+(v_a)
\prod_{a \in B}X^+(u_a):
|i;+\rangle_B \neq 0,
\end{eqnarray}
if and only if $L=n \geq 0$.

For $L=|A|-|B|+\frac{1}{2}(N-M) \geq 0$, we have the
vacuum expectation values of the $q$-vertex operators takes the form:
\begin{eqnarray}
&&
~_B\langle i;+|\Phi_{\epsilon_1}^{(i,i+1)}(\zeta_1)
\cdots
\Phi_{\epsilon_{M}}^{(i+M-1,i+M)}(\zeta_{M})
\Psi_{\mu_1}^{*(i+N,i+N-1)}(\xi_1)\cdots
\Psi_{\mu_{N}}^{*(i+1,i)}(\xi_{N})
|+;i \rangle_B
\nonumber\\
&=&(qs)^{|A|-|B|+\frac{1}{2}(N-M)}(rq^2)^{(|B|-|A|+\frac{1}{2}(M-N))i}
q^{|B|}(-q^3)^{\frac{1}{4}(M-N)^2+\frac{i}{2}(M-N)-\sum_{a \in A}a
-\sum_{b \in B}b}(1-q^2)^{|A|+|B|}\nonumber\\
&\times&
\sum_{l,m \geq 0 \atop{l+m=|A|-|B|+\frac{1}{2}(-M+N)}}
\frac{(-1)^{l+m N}
q^{\frac{-l(l+1)+m(m+1)}{2}(1-2i)-3Nm}
}{[l]_q! [m]_q!}
\prod_{j=1}^{M}
\zeta_j^{\frac{1+\epsilon_j}{2}+M-N-j+i}
\prod_{j=1}^{N} \xi_j^{\frac{1+\mu_j}{2}+N-j+1-i}
\nonumber\\
&\times&
\prod_{1\leq j<k \leq M}\frac{(q^2\zeta_k^2/\zeta_j^2;q^4)_\infty}
{(q^4\zeta_k^2/\zeta_j^2;q^4)_\infty}
\prod_{1\leq j <k \leq N}
\frac{(\xi_k^2/\xi_j^2;q^4)_\infty}{(q^2\xi_k^2/\xi_j^2;q^4)_\infty}
\prod_{j=1}^{M}\prod_{k=1}^{N} \frac{(q^3\xi_k^2/\zeta_j^2;q^4)_\infty}{
(q\xi_k^2/\zeta_j^2;q^4)_\infty}\nonumber\\
&\times& \oint \cdots \oint \prod_{a=1}^{l+m} \frac{dv_a}{2\pi \sqrt{-1}}
v_a^{2i-1}
\prod_{a \in A}\frac{dw_a}{2\pi \sqrt{-1}}w_a^{1-i}
\prod_{a \in B}\frac{du_a}{2\pi \sqrt{-1}}u_a^i
\nonumber\\
&\times&
\frac{
\prod_{a \in A}\prod_{j=1}^{N}(w_a-q^3\xi_j^2)
\prod_{a \in B}\prod_{j=1}^{M}(u_a-q^3 \zeta_j^2)
\prod_{a=1}^{l+m} \prod_{j=1}^{M}(v_a-q^3\zeta_j^2)
}{
\prod_{a \in A}
\left\{
\prod_{1\leq j \leq a}(\zeta_j^2-q^{-2}w_a)
\prod_{a \leq j \leq M}(w_a-q^4\zeta_j^2)\right\}
\prod_{a \in B}
\left\{
\prod_{1\leq j \leq a}(\xi_j^2-q^{-4}u_a)
\prod_{a \leq j \leq N}(u_a-q^2\xi_j^2)
\right\}}\nonumber\\
&\times&
\frac{
\prod_{a,b \in A
\atop{a<b}} (w_a-w_b)(w_a-q^2w_b)
\prod_{a,b \in B \atop{a<b}} (u_a-u_b)(u_a-q^{-2}u_b)
\prod_{1 \leq a<b \leq l+m}
(v_a-v_b)(v_a-q^{-2}v_b)}{
\prod_{j=1}^{N}\left\{
\prod_{a=1}^l
(v_a-q^2\xi_j^2)
\prod_{a=l+1}^{l+m}
(\xi_j^2-q^{-4}v_a)
\right\}}
\nonumber\\
&\times&
\frac{
\prod_{b \in B}\left\{
\prod_{a=1}^l (v_a-u_b)(v_a-q^{-2}u_b)
\prod_{a=l+1}^{l+m} (u_b-v_a)(u_b-q^{-2}v_a)\right\}
}{
\prod_{a \in A}\left\{
\prod_{b \in B}(w_a-qu_b)(w_a-q^{-1}u_b)
\prod_{b=1}^{l+m}
(w_a-qv_b)(w_a-q^{-1}v_b)\right\}}\nonumber
\\
&\times&
~_B\langle i|
e^{\sum_{j=1}^{M}P(\zeta_j^2)-\sum_{j=1}^{N}P(q^{-1}\xi_j^2)
+\sum_{a \in A}R^-(w_a)+\sum_{a \in B}R^+(u_a)+\sum_{a=1}^{l+m}R^+(v_a)}\nonumber\\
&\times&
e^{\sum_{j=1}^{M}Q(\zeta_j^2)-\sum_{j=1}^{N}Q(q\xi_j^2)
+\sum_{a \in A}S^-(w_a)+\sum_{a \in B}S^+(u_a)+\sum_{a=1}^{l+m}S^+(v_a)}
|i\rangle_B.
\end{eqnarray}
%%%%%%%%%%%%%%%%%%%%%%%%%%%%%%%%%%%%%%%%%%%%%%%%%%%%%%%%%%%%%%%%%
Using (\ref{eqn:vev}), we calculate the following 
vacuum expectation value :
\begin{eqnarray}
&&~_B\langle i|
e^{\sum_{j=1}^{M}P(\zeta_j^2)-\sum_{j=1}^{N}P(q^{-1}\xi_j^2)
+\sum_{a \in A}R^-(w_a)+\sum_{a \in B}R^+(u_a)+\sum_{a=1}^{L}
R^+(v_a)}\nonumber\\
&\times&
e^{\sum_{j=1}^{M}Q(\zeta_j^2)-\sum_{j=1}^{N}Q(q\xi_j^2)
+\sum_{a \in A}S^-(w_a)+\sum_{a \in B}S^+(u_a)
+\sum_{a=1}^{L}S^+(v_a)}
|i\rangle_B\nonumber\\
&=&
\langle i|e^{G^{(i)}}e^{\sum_{n=1}^\infty a_{-n}X_n}e^{-\sum_{n=1}^\infty
a_nY_n}e^{F^{(i)}}|i\rangle.
\end{eqnarray}
Here we have defined
\begin{eqnarray}
X_n&=&\frac{q^{7n/2}}{[2n]_q}\sum_{j=1}^{M}\zeta_j^{2n}
-\frac{q^{5n/2}}{[2n]_q}\sum_{j=1}^{N}\xi_j^{2n}
-\frac{q^{n/2}}{[n]_q}\sum_{a \in A}w_a^n
+\frac{q^{-n/2}}{[n]_q}\sum_{a \in B}u_a^n
+\frac{q^{-n/2}}{[n]_q}\sum_{a=1}^{L}v_a^n,\\
Y_n&=&
\frac{q^{-5n/2}}{[2n]_q}\sum_{j=1}^{M}\zeta_j^{-2n}
-\frac{q^{-7n/2}}{[2n]_q}\sum_{j=1}^{N}\xi_j^{-2n}
-\frac{q^{n/2}}{[n]_q}\sum_{a \in A}w_a^{-n}
+\frac{q^{-n/2}}{[n]_q}\sum_{a \in B}u_a^{-n}
+\frac{q^{-n/2}}{[n]_q}\sum_{a=1}^{L}v_a^{-n}.\nonumber\\
\end{eqnarray}
%where we have used $\alpha_n$, $\gamma_n$, $\beta_n^{(i)}$,
%and $\delta_n^{(i)}$ in 
%(\ref{def:alpha}), (\ref{def:beta}), and (\ref{def:delta}).
%%%%%%%%%%%%%%%%%%%%%%%%%%%%%%%%%
~\\
$\bullet$~
The followings is the main results of this Section.
For $L=|A|-|B|+\frac{1}{2}(N-M)<0$, we have
\begin{eqnarray}
F^{(+,i,M,N)}_{\epsilon_1,\cdots,\epsilon_M,\mu_1,\cdots,\mu_N}
(\zeta_1,\cdots,\zeta_M,\xi_1,\cdots,\xi_N;r,s)=0.
\end{eqnarray}
For $L=|A|-|B|+\frac{1}{2}(N-M)=0$, the form factors for triangular boundary conditions coincide with those for diagonal boundary conditions :
\begin{eqnarray}
F^{(+,i,M,N)}_{\epsilon_1,\cdots,\epsilon_M,\mu_1,\cdots,\mu_N}
(\zeta_1,\cdots,\zeta_M,\xi_1,\cdots,\xi_N;r,s)=
F^{(i,M,N)}_{\epsilon_1,\cdots,\epsilon_M,\mu_1,\cdots,\mu_N}
(\zeta_1,\cdots,\zeta_M,\xi_1,\cdots,\xi_N;r).
\end{eqnarray}
For $L=|A|-|B|+\frac{1}{2}(N-M)>0$,
 the following integral representations of the form factors
of local operators are obtained:
\begin{eqnarray}
&&
F^{(+,i,M,N)}_{\epsilon_1,\cdots,\epsilon_M,\mu_1,\cdots,\mu_N}
(\zeta_1,\cdots,\zeta_M,\xi_1,\cdots,\xi_N; r,s)
\nonumber\\
&=&
(qs)^{L}(rq^2)^{-L i}q^{|B|}(1-q^2)^{|B|}
(-q^3)^{\frac{1}{4}(M-N)^2+\frac{i}{2}(M-N)
-\sum_{a \in A}a-\sum_{b \in B}b}\nonumber\\
&\times&
\left(\frac{\{q^6\}_\infty}{\{q^8\}_\infty}\right)^{M}
\left(\frac{\{q^4\}_\infty}{\{q^6\}_\infty}\right)^{N}
\{(q^2;q^2)_\infty\}^{2|A|+\frac{1}{2}(-M+N)}
\prod_{j=1}^{M}\zeta_j^{\frac{1+\epsilon_j}{2}+M-N-j+i}
\prod_{j=1}^{N}\xi_j^{\frac{\mu_j+1}{2}+N-j+1-i}
\nonumber\\
&\times&
\prod_{1\leq j<k \leq M}\frac{
\{q^6\zeta_j^2 \zeta_k^2\}_\infty
\{q^2/\zeta_j^2\zeta_k^2\}_\infty
\{q^6\zeta_j^2/\zeta_k^2\}_\infty
\{q^2\zeta_k^2/\zeta_j^2\}_\infty
}{
\{q^8\zeta_j^2\zeta_k^2\}_\infty 
\{q^4/\zeta_j^2\zeta_k^2\}_\infty 
\{q^8\zeta_j^2/\zeta_k^2\}_\infty 
\{q^4\zeta_k^2/\zeta_j^2\}_\infty}
\nonumber\\
&\times&
\prod_{1\leq j<k \leq N}\frac{
\{q^4\xi_j^2 \xi_k^2\}_\infty
\{1/\xi_j^2\xi_k^2\}_\infty
\{q^4\xi_j^2/\xi_k^2\}_\infty
\{\xi_k^2/\xi_j^2\}_\infty
}{
\{q^6\xi_j^2\xi_k^2\}_\infty 
\{q^2/\xi_j^2\xi_k^2\}_\infty 
\{q^6\xi_j^2/\xi_k^2\}_\infty 
\{q^2\xi_k^2/\xi_j^2\}_\infty}\nonumber\\
&\times&
\prod_{j=1}^{M}
\prod_{k=1}^{N}
\frac{
\{q^7\zeta_j^2 \xi_k^2\}_\infty
\{q^3/\zeta_j^2 \xi_k^2\}_\infty
\{q^7 \zeta_j^2/\xi_k^2\}_\infty
\{q^3 \xi_k^2/\zeta_j^2\}_\infty
}{
\{q^5 \zeta_j^2\xi_k^2\}_\infty 
\{q/ \zeta_j^2\xi_k^2\}_\infty 
\{q^5 \zeta_j^2/\xi_k^2\}_\infty 
\{q \xi_k^2/\zeta_j^2\}_\infty}\nonumber\\
&\times&
\prod_{j=1}^{M} 
\frac{[q^{10}\zeta_j^4]_\infty [q^{14}\zeta_j^4]_\infty 
[q^{10}/\zeta_j^4]_\infty [q^6 /\zeta_j^4]_\infty}{
[q^{12}\zeta_j^4]_\infty [q^{16}\zeta_j^4]_\infty 
[q^{12}/\zeta_j^4]_\infty [q^8/\zeta_j^4]_\infty}
\prod_{j=1}^{N} 
\frac{[q^{4}\xi_j^4]_\infty [q^{8}\xi_j^4]_\infty 
[1/\xi_j^4]_\infty [q^4 /\xi_j^4]_\infty}{
[q^{6}\xi_j^4]_\infty [q^{10}\xi_j^4]_\infty 
[q^{2}/\xi_j^4]_\infty [q^{6}/\xi_j^4]_\infty}
\nonumber
\\
&\times&
\sum_{l,m \geq 0
\atop{l+m=L}}
\frac{(-1)^{l+mN} q^{\frac{l(l+1)-m(m+1)}{2}(2i-1)-3Nm}}
{[l]_q![m]_q!}
\oint \cdots \oint_{C^{(+,i)}_l}
\prod_{a \in A}\frac{dw_a}{2\pi \sqrt{-1}}w_a^{1-i}
\prod_{b \in B}\frac{du_b}{2\pi \sqrt{-1}}u_b^i
\prod_{c=1}^{L}\frac{dv_c}{2\pi \sqrt{-1}}v_c^{2i-1}
\nonumber\\
%%%%%%%%%%%%%%%%%%%%%%%%%%%%%%%%%%%%%%%%%%%%%%%%%%%%%%%%
&&\prod_{a \in A}(w_a^2/q^2;q^4)_\infty (q^6/w_a^2;q^4)_\infty
\prod_{a \in B}(u_a^2/q^2;q^4)_\infty (q^6/u_a^2;q^4)_\infty
\prod_{a=1}^{L}(v_a^2/q^2;q^4)_\infty 
(q^6/v_a^2;q^4)_\infty
\nonumber\\
%%%%%%%%%%%%%%%%%%%%%%%%%%%%%%%%%%%%%%%%%%%%%%%%%%%%%%
&\times&
\frac{
\prod_{1\leq a<b \leq L}
v_a^2 (v_av_b/q^4;q^2)_\infty 
(q^2v_a/v_b;q^2)_\infty (v_b/q^2v_a;q^2)_\infty 
(q^4/v_av_b;q^2)_\infty
}
{
\prod_{a \in A} \prod_{b \in B}
w_a^2 (w_au_b/q^3;q^2)_\infty 
(q^3 w_a/u_b;q^2)_\infty 
(u_b/qw_a;q^2)_\infty 
(q^5/w_a u_b;q^2)_\infty}
\nonumber\\
%%%%%%%%%%%%%%%%%%%%%%%%%%%%%%%%%%%%%%%%%%%%
&\times&
\frac{
\prod_{a=1}^L\prod_{b \in B}
(v_au_b/q^4;q^2)_\infty 
(q^2v_a/u_b;q^2)_\infty 
(q^2u_b/v_a;q^2)_\infty 
(q^4/v_au_b;q^2)_\infty
}
{
\prod_{a=1}^L \prod_{b \in A}
v_a^2 (v_aw_b/q^3;q^2)_\infty 
(q^3 v_a/w_b;q^2)_\infty 
(w_b/qv_a;q^2)_\infty 
(q^5/v_a w_b;q^2)_\infty}
\nonumber\\
%%%%%%%%%%%%%%%%%%%%%%%%%%%%%%%%%%%%%%%%%%%%
&\times&
\frac{ \prod_{a,b \in A
\atop{a<b}}w_a^2
(w_aw_b/q^2;q^2)_\infty 
(q^4w_a/w_b;q^2)_\infty 
(w_b/w_a;q^2)_\infty 
(q^6/w_aw_b;q^2)_\infty}
{
\prod_{a \in A}
\left\{\prod_{1\leq j \leq a}(\zeta_j^2-q^{-2}w_a)
\prod_{a\leq j \leq M}(w_a-q^4 \zeta_j^2)\right\}
}\nonumber\\
&\times&
\frac{
\prod_{a,b \in B
\atop{a<b}}u_a^2
(u_au_b/q^4;q^2)_\infty 
(q^2u_a/u_b;q^2)_\infty 
(u_b/q^2u_a;q^2)_\infty 
(q^4/u_au_b;q^2)_\infty}{
\prod_{a \in B}
\left\{\prod_{1\leq j \leq a}(\xi_j^2-q^{-4}u_a)
\prod_{a\leq j \leq M}(u_a-q^2 \xi_j^2)\right\}}
\nonumber\\
&\times&
\frac{
\prod_{j=1}^{M}
\prod_{a \in B} 
u_a(q\zeta_j^2 u_a;q^4)_\infty (q^3\zeta_j^2/u_a;q^4)_\infty 
(qu_a/\zeta_j^2;q^4)_\infty (q^3/\zeta_j^2 u_a;q^4)_\infty
}{
\prod_{j=1}^{M}
\prod_{a \in A}
(q^2 \zeta_j^2 w_a;q^4)_\infty 
(q^8 \zeta_j^2/w_a;q^4)_\infty (q^2w_a/\zeta_j^2;q^4)_\infty 
(q^4/\zeta_j^2 w_a;q^4)_\infty}
\nonumber
\\
%%%%%%%%%%%%%%%%%%%%%%%%%%%%%%%%%%%%%%%%%%
&\times&
\frac{
\prod_{j=1}^{N}
\prod_{a \in A} 
w_a (q\xi_j^2 w_a;q^4)_\infty 
(q^3\xi_j^2/w_a;q^4)_\infty 
(q w_a/\xi_j^2;q^4)_\infty 
(q^3/\xi_j^2 w_a;q^4)_\infty
}{
\prod_{j=1}^{N}
\prod_{a \in B}
(\xi_j^2 u_a;q^4)_\infty 
(q^6 \xi_j^2/u_a;q^4)_\infty 
(u_a/\xi_j^2;q^4)_\infty 
(q^2/\xi_j^2 u_a;q^4)_\infty}
\nonumber\\
%%%%%%%%%%%%%%%%%%%%%%%%%%%%%%%%%%%%%%%%%%%%%%%%%%%
&\times&
\frac{
\prod_{j=1}^{M}
\prod_{a=1}^L 
v_a (q\zeta_j^2 v_a;q^4)_\infty 
(q^3\zeta_j^2/v_a;q^4)_\infty 
(q v_a/\zeta_j^2;q^4)_\infty 
(q^3/\zeta_j^2 v_a;q^4)_\infty
}{
\prod_{j=1}^{N}
\prod_{a=1}^L
(\xi_j^2 v_a;q^4)_\infty 
(q^6 \xi_j^2/v_a;q^4)_\infty 
(v_a/\xi_j^2;q^4)_\infty 
(q^2/\xi_j^2 v_a;q^4)_\infty}
\nonumber
\\
&\times&
\frac{
\prod_{b \in B}
\left\{
\prod_{a=1}^l(v_a-u_b)(v_a-q^{-2}u_b)
\prod_{a=l+1}^{L}(u_b-v_a)(u_b-q^{-2}v_a)
\right\}}{
\prod_{j=1}^{N}
\left\{
\prod_{a=1}^l(v_a-q^2\xi_j^2)
\prod_{a=l+1}^L(\xi_j^2-q^{-4}v_a)
\right\}}
\nonumber
\\
&\times&
\left\{\begin{array}{cc}
\prod_{j=1}^{M}\frac{(q^2r\zeta_j^2;q^4)_\infty}{(q^4r\zeta_j^2;q^4)_\infty}
\prod_{j=1}^{N}\frac{(q^3r \xi_j^2;q^4)_\infty}{
(qr\xi_j^2;q^4)_\infty}
\frac{ \prod_{a \in B}(1-ru_a/q^3)
\prod_{a=1}^L (1-rv_a/q^3)
}{
 \prod_{a \in A}(1-rw_a/q^2)}& (i=0),\\
\prod_{j=1}^{M}\frac{(1/r\zeta_j^2;q^4)_\infty}{(q^2/r\zeta_j^2;q^4)_\infty}
\prod_{j=1}^{N}
\frac{(q/r\xi_j^2;q^4)_\infty}{(1/qr\xi_j^2;q^4)_\infty}
\frac{ 
\prod_{a \in B}(1-q/ru_a)
\prod_{a=1}^L(1-q/rv_a)
}{
\prod_{a \in A}(1-q^2/rw_a)}& (i=1).
\end{array}\right.
\end{eqnarray}
Here the integration contour $C^{(+,0)}_l~(1\leq l \leq L)$ is a closed curve that satisfies 
the following conditions for $s=0,1,2,\cdots$.
The $w_a~(a\in A)$ encircles
$q^{8+4s}\zeta_j^2~(1\leq j <a)$,
$q^{4+4s}\zeta_j^2~(a \leq j \leq M)$,
$q^{4+4s}/\zeta_j^2~(1\leq j \leq M)$,
$q^{-1+2s}u_b,~ q^{5+2s}/u_b~(b \in B)$,
$q^{3+2s}v_c~(1\leq c \leq l)$,
$q^{-1+2s}v_c~(l<c \leq L)$,
$q^{5+2s}/v_c~(1\leq c \leq L)$,
but not
$q^{2-4s}\zeta_j^2~(1\leq j \leq a)$,
$q^{-2-4s}\zeta_j^2~(a < j \leq M)$,
$q^{-2-4s}/\zeta_j^2~(1\leq j \leq M)$,
$q^{-3-2s}u_b,~ q^{3-2s}/u_b~(b \in B)$, $q^2/r$,
$q^{1-2s}v_c~(1\leq c \leq l)$,
$q^{-3-2s}v_c~(l<c \leq L)$,
$q^{3-2s}/v_c~(1\leq c \leq L)$.
The $u_b~(b \in B)$ encircles
$q^{6+4s}\xi_j^2~(1\leq j <b)$,
$q^{2+4s}\xi_j^2~(b \leq j \leq N)$,
$q^{2+4s}/\xi_j^2~(1\leq j \leq N)$,
$q^{3+2s}w_a,~ q^{5+2s}/w_a~(a \in B)$,
but not
$q^{4-4s}\xi_j^2~(1\leq j \leq b)$,
$q^{-4s}\xi_j^2~(b < j \leq N)$,
$q^{-4s}/\xi_j^2~(1\leq j \leq N)$,
$q^{1-2s}w_a,~ q^{-3-2s}/w_a~(a \in A)$.
The $v_c~(1\leq c \leq l)$ encircles
$q^{2+4s}\xi_j^2,~q^{2+4s}/\xi_j^2~(1\leq j \leq N)$,
$q^{-1+2s}w_a,~q^{5+2s}/w_a~(a \in A)$,
but not
$q^{-4s}\xi_j^2,~q^{-4s}/\xi_j^2~(1\leq j \leq N)$,
$q^{-3-2s}w_a,~q^{3-2s}/w_a~(a\in A)$.
The $v_c~(l< c \leq L)$ encircles
$q^{6+4s}\xi_j^2,~q^{2+4s}/\xi_j^2~(1\leq j \leq N)$,
$q^{3+2s}w_a,~q^{5+2s}/w_a~(a\in A)$,
but not
$q^{4-4s}\xi_j^2,~q^{-4s}/\xi_j^2~(1\leq j \leq N)$,
$q^{1-2s}w_a,~q^{3-2s}/w_a~(a\in A)$.
Here the integration contour $C^{(+,1)}_l~(1\leq l \leq L)$ is a closed curve such that $w_a~(a \in A)$ encircles
$q^2/r$ in addition the same points $C^{(+,0)}_l$ does.

\subsection{Second integral representation}

In this Section we focus our attention on
the half-infinite XXZ spin chain with a lower-triangular boundary.
Let us consider the form factors with $M=N~(mod.2)$ : 
\begin{eqnarray}
&&F^{(-,i,M,N)}_{\epsilon_1,\cdots,\epsilon_M,\mu_1,\cdots,\mu_N}(\zeta_1,\cdots,\zeta_M,\xi_1,\cdots,\xi_N;r,s).
\end{eqnarray}
In what follows we use
$A=\{1\leq a \leq M|\epsilon_a=+\}$, $B=\{1\leq b \leq N|\mu_b=+\}$,
and 
\begin{eqnarray}
K=|B|-|A|+\frac{1}{2}(M-N).
\end{eqnarray}
In the following, we understand $K=0$ as a sufficient condition
such that the expressions for triangular boundary conditions coincide with 
those for diagonal boundary conditions. 
The following bosonizations of the Chevalley generators are used:
\begin{eqnarray}
f_1=\oint \frac{dv}{2\pi \sqrt{-1}}X^-(v),~~~
e_0q^{-h_0}=q^{-1} \oint \frac{dv}{2\pi \sqrt{-1}}vX^+(v).
\end{eqnarray}
The bosonic parts of the following operators :
\begin{eqnarray}
&&
\exp_{q^{-1}}\left(
\frac{s}{q}e_0 q^{-h_0}\right)
\Phi_{\epsilon_1}^{(0,1)}(\zeta_1)
\cdots
\Phi_{\epsilon_{M}}^{(M-1,M)}(\zeta_{M})
\Psi_{\mu_1}^{*(N,N-1)}(\xi_1)\cdots
\Psi_{\mu_{N}}^{*(1,0)}(\xi_{N})
\exp_{q}\left(
-\frac{s}{q}e_0 q^{-h_0}\right),\nonumber
\\
&&
\exp_{q}\left(
\frac{s}{r}f_1\right)
\Phi_{\epsilon_1}^{(1,0)}(\zeta_1)
\cdots
\Phi_{\epsilon_{M}}^{(M,M+1)}(\zeta_{M})
\Psi_{\mu_1}^{*(N+1,N)}(\xi_1)\cdots
\Psi_{\mu_{2N}}^{*(0,1)}(\xi_{N})
\exp_{q^{-1}}\left(
-\frac{s}{r}f_1\right),
\end{eqnarray}
are given by
\begin{eqnarray}
:\prod_{j=1}^M
\Phi_{-}^{(i+j-1,i+j)}(\zeta_j)
\prod_{j=1}^N
\Psi_{-}^{*(i+N-j+1,i+N-j)}(\xi_j)
\prod_{a \in A}X^-(w_a)
\prod_{a \in B}X^+(u_a)
\prod_{a=1}^{n}X^-(v_a):,
\end{eqnarray}
for $n=0,1,2,\cdots$.
Here the parameter $n$ comes from the $q$-exponential.
The zero-mode of the above operator is 
$e^{\alpha(-n-|A|+|B|+\frac{1}{2}(M-N))}$,
hence the vacuum expectation value becomes non-zero :
\begin{eqnarray}
~_B\langle-;i|:\prod_{j=1}^M
\Phi_{-}^{(i+j-1,i+j)}(\zeta_j)
\prod_{j=1}^N
\Psi_{-}^{*(i+N-j+1,i+N-j)}(\xi_j)
\prod_{a \in A}X^-(w_a)
\prod_{a \in B}X^+(u_a)
\prod_{a=1}^{n}X^-(v_a):|-;i\rangle_B\neq 0,\nonumber\\
\end{eqnarray}
if and only if $K=n\geq 0$.

For $K=|B|-|A|+\frac{1}{2}(M-N)\geq 0$, the vacuum expectation values of the $q$-vertex operators 
read:
\begin{eqnarray}
&&
~_B\langle i;-|
\Phi_{\epsilon_1}^{(i,i+1)}(\zeta_1)
\cdots
\Phi_{\epsilon_{M}}^{(i+M-1,i+M)}(\zeta_{M})
\Psi_{\mu_1}^{*(i+N,i+N-1)}(\xi_1)\cdots
\Psi_{\mu_{N}}^{*(1-i,i)}(\xi_{N})
|-;i\rangle_B
\nonumber\\
&=&
(s/q^2)^{\frac{1}{2}(M-N)-|A|+|B|}
(q^2/r)^{(\frac{1}{2}(M-N)-|A|+|B|)i}
q^{|B|}(-q^3)^{\frac{1}{4}(M-N)^2+\frac{i}{2}(M-N)-\sum_{a \in A}a
-\sum_{b \in B}b}
(1-q^2)^{|A|+|B|}\nonumber\\
&\times&
\sum_{l,m \geq 0 \atop{l+m=\frac{1}{2}(M-N)-|A|+|B|}}
\frac{(-1)^{m(M+1)}
q^{\frac{l(l-1)-m(m-1)}{2}(2i-1)-3Mm}
}{[l]_q! [m]_q!}
\prod_{j=1}^{M}
\zeta_j^{\frac{1+\epsilon_j}{2}+M-N-j+i}
\prod_{j=1}^{N} \xi_j^{\frac{1+\mu_j}{2}+N-j+1-i}
\nonumber\\
&\times&
\prod_{1\leq j<k \leq M}\frac{(q^2\zeta_k^2/\zeta_j^2;q^4)_\infty}
{(q^4\zeta_k^2/\zeta_j^2;q^4)_\infty}
\prod_{1\leq j <k \leq N}
\frac{(\xi_k^2/\xi_j^2;q^4)_\infty}{(q^2\xi_k^2/\xi_j^2;q^4)_\infty}
\prod_{j=1}^{M}\prod_{k=1}^{N} \frac{(q^3\xi_k^2/\zeta_j^2;q^4)_\infty}{
(q\xi_k^2/\zeta_j^2;q^4)_\infty}\nonumber\\
&\times& \oint \cdots \oint \prod_{a=1}^{l+m} 
\frac{dv_a}{2\pi \sqrt{-1}}v_a^{1-2i}
\prod_{a \in A}\frac{dw_a}{2\pi \sqrt{-1}}w_a^{1-i}
\prod_{a \in B}\frac{du_a}{2\pi \sqrt{-1}}u_a^i
\nonumber\\
&\times&
\frac{
\prod_{a \in A}\prod_{j=1}^{N}(w_a-q^3\xi_j^2)
\prod_{a \in B}\prod_{j=1}^{M}(u_a-q^3 \zeta_j^2)
\prod_{a=1}^{l+m} \prod_{j=1}^{N}(v_a-q^3\xi_j^2)
}{
\prod_{a \in A}
\left\{
\prod_{1\leq j \leq a}(\zeta_j^2-q^{-2}w_a)
\prod_{a \leq j \leq M}(w_a-q^4\zeta_j^2)\right\}
\prod_{a \in B}
\left\{
\prod_{1\leq j \leq a}(\xi_j^2-q^{-4}u_a)
\prod_{a \leq j \leq N}(u_a-q^2\xi_j^2)
\right\}}\nonumber\\
&\times&
\frac{
\prod_{a,b \in A
\atop{a<b}} (w_a-w_b)(w_a-q^2w_b)
\prod_{a,b \in B \atop{a<b}} (u_a-u_b)(u_a-q^{-2}u_b)
\prod_{1 \leq a<b \leq l+m}
(v_a-v_b)(v_a-q^{2}v_b)}{
\prod_{j=1}^{M}\left\{
\prod_{a=1}^l
(v_a-q^4\zeta_j^2)
\prod_{a=l+1}^{l+m}
(\zeta_j^2-q^{-2}v_a)
\right\}}
\nonumber\\
&\times&
\frac{
\prod_{a \in A}\left\{
\prod_{b=1}^l (v_b-w_a)(v_b-q^{2}w_a)
\prod_{b=l+1}^{l+m} (w_a-v_b)(w_a-q^{2}v_b)\right\}
}{
\prod_{b \in B}\left\{
\prod_{a \in A}(w_a-qu_b)(w_a-q^{-1}u_b)
\prod_{a=1}^{l+m}
(v_a-qu_b)(v_a-q^{-1}u_b)\right\}}\nonumber
\\
&\times&
~_B\langle i|
e^{\sum_{j=1}^{M}P(\zeta_j^2)-\sum_{j=1}^{N}P(q^{-1}\xi_j^2)
+\sum_{a \in A}R^-(w_a)+\sum_{a \in B}R^+(u_a)
+\sum_{a=1}^{l+m}R^-(v_a)}\nonumber\\
&\times&
e^{\sum_{j=1}^{M}Q(\zeta_j^2)-\sum_{j=1}^{N}Q(q\xi_j^2)
+\sum_{a \in A}S^-(w_a)+\sum_{a \in B}S^+(u_a)+\sum_{a=1}^{l+m}S^-(v_a)}
|i\rangle_B.
\end{eqnarray}
Using (\ref{eqn:vev}),
we calculate the following vacuum expectation value :
\begin{eqnarray}
&&~_B\langle i|
e^{\sum_{j=1}^{M}P(\zeta_j^2)-\sum_{j=1}^{N}P(q^{-1}\xi_j^2)
+\sum_{a \in A}R^-(w_a)+\sum_{a \in B}R^+(u_a)+\sum_{a=1}^{l+m}R^-(v_a)}\nonumber\\
&\times&
e^{\sum_{j=1}^{M}Q(\zeta_j^2)-\sum_{j=1}^{N}Q(q\xi_j^2)
+\sum_{a \in A}S^-(w_a)+\sum_{a \in B}S^+(u_a)+\sum_{a=1}^{l+m}S^-(v_a)}
|i\rangle_B\nonumber\\
&=&
\langle i|e^{G^{(i)}}e^{\sum_{n=1}^\infty a_{-n}X_n}e^{-\sum_{n=1}^\infty
a_nY_n}e^{F^{(i)}}|i\rangle.
\end{eqnarray}
Here we have defined
\begin{eqnarray}
X_n&=&\frac{q^{7n/2}}{[2n]_q}\sum_{j=1}^{M}\zeta_j^{2n}
-\frac{q^{5n/2}}{[2n]_q}\sum_{j=1}^{N}\xi_j^{2n}
-\frac{q^{n/2}}{[n]_q}\sum_{a \in A}w_a^n
+\frac{q^{-n/2}}{[n]_q}\sum_{a \in B}u_a^n
-\frac{q^{n/2}}{[n]_q}\sum_{a=1}^{K}v_a^n,\\
Y_n&=&
\frac{q^{-5n/2}}{[2n]_q}\sum_{j=1}^{M}\zeta_j^{-2n}
-\frac{q^{-7n/2}}{[2n]_q}\sum_{j=1}^{N}\xi_j^{-2n}
-\frac{q^{n/2}}{[n]_q}\sum_{a \in A}w_a^{-n}
+\frac{q^{-n/2}}{[n]_q}\sum_{a \in B}u_a^{-n}
-\frac{q^{n/2}}{[n]_q}\sum_{a=1}^{K}v_a^{-n}.\nonumber\\
\end{eqnarray}

~\\
$\bullet$~The followings are the main results of this Section.
For $K=|B|-|A|+\frac{1}{2}(M-N)<0$
we have
\begin{eqnarray}
F^{(-,i,M,N)}_{\epsilon_1,\cdots,\epsilon_M,\mu_1,\cdots,\mu_N}
(\zeta_1,\cdots,\zeta_M,\xi_1,\cdots,\xi_N;r,s)=0.
\end{eqnarray}
For $K=|B|-|A|+\frac{1}{2}(M-N)=0$, 
note that the form factors for triangular boundary conditions 
coincide with those for diagonal boundary conditions :
\begin{eqnarray}
F^{(-,i,M,N)}_{\epsilon_1,\cdots,\epsilon_M,\mu_1,\cdots,\mu_N}
(\zeta_1,\cdots,\zeta_M,\xi_1,\cdots,\xi_N;r,s)=
F^{(i,M,N)}_{\epsilon_1,\cdots,\epsilon_M,\mu_1,\cdots,\mu_N}
(\zeta_1,\cdots,\zeta_M,\xi_1,\cdots,\xi_N;r).
\end{eqnarray}
For $K=|B|-|A|+\frac{1}{2}(M-N)>0$,
the following integral representations of the form factors
of local operators is obtained:
\begin{eqnarray}
&&
F^{(-,i,M,N)}_{\epsilon_1,\cdots,\epsilon_M,\mu_1,\cdots,\mu_N}
(\zeta_1,\cdots,\zeta_M,\xi_1,\cdots,\xi_N;r,s)
\nonumber\\
&=&
(s/q^2)^K (q^2/r)^{i K}
q^{|B|}(1-q^2)^{|A|+2|B|}
(-q^3)^{\frac{1}{4}(M-N)^2+\frac{i}{2}(M-N)
-\sum_{a \in A}a-\sum_{b \in B}b}\nonumber\\
&\times&
\left(\frac{\{q^6\}_\infty}{\{q^8\}_\infty}\right)^{M}
\left(\frac{\{q^4\}_\infty}{\{q^6\}_\infty}\right)^{N}
\{(q^4;q^2)_\infty\}^{2|B|+\frac{1}{2}(M-N)}
\prod_{j=1}^{M}\zeta_j^{\frac{1+\epsilon_j}{2}+M-N-j+i}
\prod_{j=1}^{N}\xi_j^{\frac{\mu_j+1}{2}+N-j+1-i}
\nonumber\\
&\times&
\prod_{1\leq j<k \leq M}\frac{
\{q^6\zeta_j^2 \zeta_k^2\}_\infty
\{q^2/\zeta_j^2\zeta_k^2\}_\infty
\{q^6\zeta_j^2/\zeta_k^2\}_\infty
\{q^2\zeta_k^2/\zeta_j^2\}_\infty
}{
\{q^8\zeta_j^2\zeta_k^2\}_\infty 
\{q^4/\zeta_j^2\zeta_k^2\}_\infty 
\{q^8\zeta_j^2/\zeta_k^2\}_\infty 
\{q^4\zeta_k^2/\zeta_j^2\}_\infty}
\nonumber\\
&\times&
\prod_{1\leq j<k \leq N}\frac{
\{q^4\xi_j^2 \xi_k^2\}_\infty
\{1/\xi_j^2\xi_k^2\}_\infty
\{q^4\xi_j^2/\xi_k^2\}_\infty
\{\xi_k^2/\xi_j^2\}_\infty
}{
\{q^6\xi_j^2\xi_k^2\}_\infty 
\{q^2/\xi_j^2\xi_k^2\}_\infty 
\{q^6\xi_j^2/\xi_k^2\}_\infty 
\{q^2\xi_k^2/\xi_j^2\}_\infty}\nonumber\\
&\times&
\prod_{j=1}^{M}
\prod_{k=1}^{N}
\frac{
\{q^7\zeta_j^2 \xi_k^2\}_\infty
\{q^3/\zeta_j^2 \xi_k^2\}_\infty
\{q^7 \zeta_j^2/\xi_k^2\}_\infty
\{q^3 \xi_k^2/\zeta_j^2\}_\infty
}{
\{q^5 \zeta_j^2\xi_k^2\}_\infty 
\{q/ \zeta_j^2\xi_k^2\}_\infty 
\{q^5 \zeta_j^2/\xi_k^2\}_\infty 
\{q \xi_k^2/\zeta_j^2\}_\infty}\nonumber\\
&\times&
\prod_{j=1}^{M} 
\frac{[q^{10}\zeta_j^4]_\infty [q^{14}\zeta_j^4]_\infty 
[q^{10}/\zeta_j^4]_\infty [q^6 /\zeta_j^4]_\infty}{
[q^{12}\zeta_j^4]_\infty [q^{16}\zeta_j^4]_\infty 
[q^{12}/\zeta_j^4]_\infty [q^8/\zeta_j^4]_\infty}
\prod_{j=1}^{N} 
\frac{[q^{4}\xi_j^4]_\infty [q^{8}\xi_j^4]_\infty 
[1/\xi_j^4]_\infty [q^4 /\xi_j^4]_\infty}{
[q^{6}\xi_j^4]_\infty [q^{10}\xi_j^4]_\infty 
[q^{2}/\xi_j^4]_\infty [q^{6}/\xi_j^4]_\infty}
\nonumber
\\
&\times&
\sum_{l,m \geq 0 \atop{l+m=K}}
\frac{(-1)^{m(M+1)} q^{\frac{l(l-1)-m(m-1)}{2}(2i-1)-3Mm}}{[l]_q![m]_q!}
\oint \cdots \oint_{C^{(-,i)}_l}
\prod_{a \in A}\frac{dw_a}{2\pi \sqrt{-1}}w_a^{1-i}
\prod_{b \in B}\frac{du_b}{2\pi \sqrt{-1}}u_b^i
\prod_{c=1}^K \frac{dv_c}{2\pi \sqrt{-1}}v_c^{1-2i}
\nonumber\\
%%%%%%%%%%%%%%%%%%%%%%%%%%%%%%%%%%%%%%%%%%%%%%%%%%%%%%%%%%
&\times&
\prod_{a \in A}(w_a^2/q^2;q^4)_\infty (q^6/w_a^2;q^4)_\infty
\prod_{a \in B}(u_a^2/q^2;q^4)_\infty (q^6/u_a^2;q^4)_\infty
\prod_{a=1}^K (v_a^2/q^2;q^4)_\infty (q^6/v_a^2;q^4)_\infty
\nonumber\\
&\times&
\frac{
\prod_{1\leq a<b \leq K}v_a^2
(v_av_b/q^2;q^2)_\infty 
(q^4 v_a/v_b;q^2)_\infty 
(v_b/v_a;q^2)_\infty 
(q^6/v_a v_b;q^2)_\infty
}
{
\prod_{a \in A} \prod_{b \in B}
w_a^2 (w_au_b/q^3;q^2)_\infty 
(q^3 w_a/u_b;q^2)_\infty 
(u_b/qw_a;q^2)_\infty 
(q^5/w_a u_b;q^2)_\infty}
\nonumber\\
&\times&
\frac{
\prod_{a=1}^K \prod_{b \in A}
(v_aw_b/q^2;q^2)_\infty 
(q^4 v_a/w_b;q^2)_\infty 
(w_b/v_a;q^2)_\infty 
(q^6/v_a w_b;q^2)_\infty
}
{
\prod_{a=1}^K \prod_{b \in B}
v_a^2 (v_au_b/q^2;q^2)_\infty 
(q^3 v_a/u_b;q^2)_\infty 
(u_b/qv_a;q^2)_\infty 
(q^5/v_a u_b;q^2)_\infty}
\nonumber\\
%%%%%%%%%%%%%%%%%%%%%%%%%%%%%%%%%%%%%%%%%%%%%%%%%%%%%%%%%%
&\times&
\frac{ \prod_{a,b \in A
\atop{a<b}}w_a^2
(w_aw_b/q^2;q^2)_\infty 
(q^4w_a/w_b;q^2)_\infty 
(w_b/w_a;q^2)_\infty 
(q^6/w_aw_b;q^2)_\infty}
{
\prod_{a \in A}
\left\{\prod_{1\leq j \leq a}(\zeta_j^2-q^{-2}w_a)
\prod_{a\leq j \leq M}(w_a-q^4 \zeta_j^2)\right\}
}\nonumber\\
&\times&
\frac{
\prod_{a,b \in B
\atop{a<b}}u_a^2
(u_au_b/q^4;q^2)_\infty 
(q^2u_a/u_b;q^2)_\infty 
(u_b/q^2u_a;q^2)_\infty 
(q^4/u_au_b;q^2)_\infty}{
\prod_{a \in B}
\left\{\prod_{1\leq j \leq a}(\xi_j^2-q^{-4}u_a)
\prod_{a\leq j \leq M}(u_a-q^2 \xi_j^2)\right\}}
\nonumber\\
&\times&
\frac{
\prod_{j=1}^{M}
\prod_{a \in B} 
u_a(q\zeta_j^2 u_a;q^4)_\infty (q^3\zeta_j^2/u_a;q^4)_\infty 
(qu_a/\zeta_j^2;q^4)_\infty (q^3/\zeta_j^2 u_a;q^4)_\infty
}{
\prod_{j=1}^{M}
\prod_{a \in A}
(q^2 \zeta_j^2 w_a;q^4)_\infty 
(q^8 \zeta_j^2/w_a;q^4)_\infty (q^2w_a/\zeta_j^2;q^4)_\infty 
(q^4/\zeta_j^2 w_a;q^4)_\infty}
\nonumber
\\
&\times&
\frac{
\prod_{j=1}^{N}
\prod_{a=1}^K 
v_a (q\xi_j^2 v_a;q^4)_\infty 
(q^3\xi_j^2/v_a;q^4)_\infty 
(q v_a/\xi_j^2;q^4)_\infty 
(q^3/\xi_j^2 v_a;q^4)_\infty
}{
\prod_{j=1}^{M}
\prod_{a=1}^K
(q^2\zeta_j^2 v_a;q^4)_\infty 
(q^8 \xi_j^2/v_a;q^4)_\infty 
(q^2 v_a/\xi_j^2;q^4)_\infty 
(q^4/\xi_j^2 v_a;q^4)_\infty}
\nonumber\\
&\times&
\frac{
\prod_{a \in A}\left\{
\prod_{b=1}^l(v_b-w_a)(v_b-q^2w_a)
\prod_{b=l+1}^K(w_a-v_b)(w_a-q^2v_b)
\right\}}{
\prod_{a=1}^{M}\left\{
\prod_{a=1}^l(v_a-q^4\zeta_j^2)
\prod_{a=l+1}^K
(\zeta_j^2-q^{-2}v_a)\right\}}
\nonumber\\
&\times&
\left\{\begin{array}{cc}
\prod_{j=1}^{M}\frac{(q^2r\zeta_j^2;q^4)_\infty}{(q^4r\zeta_j^2;q^4)_\infty}
\prod_{j=1}^{N}\frac{(q^3r \xi_j^2;q^4)_\infty}{
(qr\xi_j^2;q^4)_\infty}
\frac{ \prod_{a \in B}(1-ru_a/q^3)}{
 \prod_{a \in A}(1-rw_a/q^2)
\prod_{a=1}^K (1-rv_a/q^2)}& (i=0),\\
\prod_{j=1}^{M}\frac{(1/r\zeta_j^2;q^4)_\infty}{(q^2/r\zeta_j^2;q^4)_\infty}
\prod_{j=1}^{N}
\frac{(q/r\xi_j^2;q^4)_\infty}{(1/qr\xi_j^2;q^4)_\infty}
\frac{ 
\prod_{a \in B}(1-q/ru_a)}{
\prod_{a \in A}(1-q^2/rw_a)
\prod_{a=1}^K(1-q^2/rv_a)}& (i=1).
\end{array}\right.
\end{eqnarray}
Here the integration contour $C^{(-,0)}_l$ is a closed curve that satisfies 
the following conditions for $s=0,1,2,\cdots$.
The $w_a~(a\in A)$ encircles
$q^{8+4s}\zeta_j^2~(1\leq j <a)$,
$q^{4+4s}\zeta_j^2~(a \leq j \leq M)$,
$q^{4+4s}/\zeta_j^2~(1\leq j \leq M)$,
$q^{-1+2s}u_b,~ q^{5+2s}/u_b~(b \in B)$,
but not
$q^{2-4s}\zeta_j^2~(1\leq j \leq a)$,
$q^{-2-4s}\zeta_j^2~(a < j \leq M)$,
$q^{-2-4s}/\zeta_j^2~(1\leq j \leq M)$,
$q^{-3-2s}u_b,~ q^{3-2s}/u_b~(b \in B)$, $q^2/r$.
The $u_b~(b \in B)$ encircles
$q^{6+4s}\xi_j^2~(1\leq j <b)$,
$q^{2+4s}\xi_j^2~(b \leq j \leq N)$,
$q^{2+4s}/\xi_j^2~(1\leq j \leq N)$,
$q^{3+2s}w_a,~ q^{5+2s}/w_a~(a \in B)$,
$q^{3+2s}v_c~(1\leq c \leq l)$,
$q^{-1+2s}v_c~(l<c\leq K)$,
$q^{5+2s}/v_c~(1\leq c \leq K)$,
but not
$q^{4-4s}\xi_j^2~(1\leq j \leq b)$,
$q^{-4s}\xi_j^2~(b < j \leq N)$,
$q^{-4s}/\xi_j^2~(1\leq j \leq N)$,
$q^{1-2s}w_a,~ q^{-3-2s}/w_a~(a \in A)$,
$q^{1-2s}v_c~(1\leq c \leq l)$,
$q^{-3-2s}v_c~(l<c \leq K)$,
$q^{3-2s}/v_c~(1\leq c \leq K)$.
The $v_c~(1\leq c \leq l)$ encircles
$q^{4+4s}\zeta_j^2,~q^{4+4s}/\zeta_j^2~(1\leq j \leq M)$,
$q^{-1+2s}u_b,~q^{5+2s}u_b~(b\in B)$,
but not
$q^{-2-4s}\zeta_j^2,~q^{-2-4s}/\zeta_j^2~(1\leq j \leq M)$,
$q^{-3-2s}u_b,~q^{3-2s}u_b~(b \in B)$, $q^2/r$.
The $v_c~(l< c \leq K)$ encircles
$q^{8+4s}\zeta_j^2,~q^{4+4s}/\zeta_j^2~(1\leq j \leq M)$,
$q^{3+2s}u_b,~q^{5+2s}u_b~(b\in B)$,
but not
$q^{2-4s}\zeta_j^2,~q^{-2-4s}/\zeta_j^2~(1\leq j \leq M)$,
$q^{1-2s}u_b,~q^{3-2s}u_b~(b \in B)$, $q^2/r$.
Here the integration contour $C^{(-,0)}_l$ is a closed curve such that $w_a~(a \in A)$ encircles
$q^2/r$ in addition the same points $C^{(-,0)}_l$ does.

\subsection{Identities between multiple integrals}

From the spin-reversal properties in (\ref{eqn:reversal-VO}), 
(\ref{eqn:reversal-vacuum-triangular}), and 
(\ref{eqn:reversal-dual-vacuum-triangular}),
we have the following relations between form factors
of local operators :
\begin{eqnarray}
&&
F^{(\pm,i,M,N)}_{\epsilon_1,\cdots,\epsilon_M,\mu_1,\cdots,\mu_N}
(\zeta_1,\cdots,\zeta_M,\xi_1,\cdots,\xi_N;r,s)\nonumber\\
&=&
F^{(\mp,1-i,M,N)}_{-\epsilon_1,\cdots,-\epsilon_M,-\mu_1,\cdots,-\mu_N}
(\zeta_1,\cdots,\zeta_M,\xi_1,\cdots,\xi_N;1/r,-s/r).
\label{eqn:reversal-triangular-form}
\end{eqnarray}
We should understand this spin-reversal property
(\ref{eqn:reversal-triangular-form}) as an analytic continuation of 
the parameter $r$.
From the above (\ref{eqn:reversal-triangular-form}), similarly to the diagonal case
we obtain infinitely many relations between $n$-fold integrals
which cannot be reduced to the relations
between $n$-fold integrals of elliptic gamma functions 
summarized in \cite{Rains}.
Here we focus on a simple example.
\\
$\bullet$~From the following relation between form factors :
\begin{eqnarray}
F^{(+,1,2,0)}_{+,+}(-q^{-1}\zeta,\zeta;r,s)=
F^{(-,0,2,0)}_{-,-}(-q^{-1}\zeta,\zeta;1/r,-s/r),
\end{eqnarray}
we deduce the identity:
\begin{eqnarray}
&&
2+\frac{1-\zeta^2/r}{\zeta^2}
\sum_{k=1}^\infty
(-q^2)^{k}\frac{
(\zeta^2-\zeta^{-2})-(1+q^{4k})/r+(\zeta^2+\zeta^{-2})q^{2k}
}{(1-q^{2k}\zeta^2/r)(1-q^{2k}/r\zeta^2)}
\nonumber\\
&=&q^2
\frac{(q^2;q^2)_\infty^8}{(q^4;q^4)_\infty^4}
\frac{\Theta_{q^4}(\zeta^4)}{1-\zeta^4}
\left(q^2 \oint \oint \oint_{C_0^{(+,1)}}
-\oint \oint \oint_{C_1^{(+,1)}}\right)
\prod_{a=1}^3 \frac{dw_a}{2\pi \sqrt{-1}}
\nonumber\\
&\times&
\frac{
(1-1/r\zeta^2)(1-q/rw_3)\prod_{a=1}^2(1-q^2/\zeta^2w_a)}{
w_2^2 w_3^3 (1-q^2w_1/w_2)(1-q^4/w_1w_2)
\prod_{a=1,2}(1-q^2/rw_a)
}\nonumber\\
&\times&
\frac{
\Theta_{q^2}(w_1w_2)
\Theta_{q^2}(w_2/w_1)
\Theta_{q^2}(\zeta^2 w_3/q)
\Theta_{q^2}(qw_3/\zeta^2)
\prod_{a=1}^3
\Theta_{q^4}(w_a^2/q^2)}{
\prod_{a=1}^2
\Theta_{q^2}(w_aw_3/q^2)
\Theta_{q^2}(w_a/qw_3)
\Theta_{q^2}(w_a\zeta^2)\Theta_{q^2}(w_a/\zeta^2)}.
\end{eqnarray}
The integration contour $C_0^{(+,1)}$ is a simple closed curve such that
the $w_1$ encircles
$q^{2+2s}\zeta^2$, $q^{4+2s}/\zeta^2$, $q^{-1+2s}w_3$,
$q^{5+2s}/w_3$, $q^2/r$
the $w_2$ encircles $q^{4+2s}\zeta^2$, $q^{4+2s}/\zeta^2$, 
$q^{-1+2s}w_3$, $q^{5+2s}/w_3$,
$q^2/r$, and the $w_3$ encircles $q^{3+2s}w_a, q^{5+2s}/w_a~(a=1,2)$,
for $s=0,1,2,\cdots$.
The integration contour $C_1^{(+,1)}$ is a simple closed curve such that
the $w_1$ encircles $q^{2+2s}\zeta^2$, 
$q^{4+2s}/\zeta^2$, $q^{3+2s}w_3$, $q^{5+2s}/w_3$,
the $w_2$ encircles $q^{4+2s}\zeta^2$, $q^{4+2s}/\zeta^2$, 
$q^{3+2s}w_3$, $q^{5+2s}/w_3$,
and $w_3$ encircles $q^{-1+2s}w_a, q^{5+2s}/w_a$ $(a=1,2)$,
for $s=0,1,2,\cdots$.

\section{Concluding remarks}
In the present article, integral representations of
form factors of local operators of the half-infinite XXZ spin chain for diagonal or triangular boundary conditions  in the massive regime have been proposed. Completing the analysis of \cite{JKKKM,BB1,BK1}, they are written in the form of multiple integrals of meromorphic functions. For some simple relations between the number of type I and type II vertex operators inserted, the expressions for diagonal or triangular boundary conditions are identical.  For diagonal boundary conditions, the expressions here obtained essentially coincide with those derived from the QISM \cite{KZMNST, KZMNST2}. In addition, the constructive approach introduced in \cite{BK1} which allows to derive identities between $n$-fold multiple integrals of elliptic theta functions (obtained from expectation values of type I vertex operators) has been extended to the family of multiple integrals associated with form factors: corresponding  identities follow from expectation values of combinations of type I and type II vertex operators. A direct proof of such remarkable identities would be highly desirable.

For more general integrable boundary conditions, the explicit construction of the vacuum eigenvectors and dual ones starting from the representation theory of the $q$-Onsager algebra can be considered following \cite{BB1}. This is currently investigated. As soon as the eigenvectors and duals will be identified, the VOA to the computation of correlation functions and form factors  can be applied in a straightforward manner, extending the results of \cite{JKKKM,BK1} and the ones here presented.  

For models with higher symmetries
\cite{Reshetikhin, Wheeler, Koyama, JKK, Bernard, KSU}, 
recall that the VOA approach has been applied successfully for diagonal boundary conditions in
some cases \cite{FK,Kojima1,Kojima2}. For the case of models associated with higher symmetries and more general boundary conditions, the analysis presented in \cite{BB1,BK1} and here may be considered. To this end, let us mention that the non-Abelian infinite dimensional algebra which occurs in the thermodynamic limit is a coideal subalgebra of the basic quantum group. For the family of affine Lie algebras $\widehat{g}$, coideal subalgebras of $U_q(\widehat{g})$ generate the so-called generalized $q$-Onsager algebras introduced in \cite{BB2}. As a starting point, infinite dimensional ($q$-vertex operators) representations have to be considered, following \cite{currents,BB1}. 

To conclude, we would like to stress that alternative derivations of the expressions here proposed would be highly desirable. Based on the recent results for a triangular boundary \cite{PL}, this may be feasible using the QISM. For more general integrable boundary conditions, other approaches than the ones based on Onsager's formulation/VOA would be required: for instance, a modified algebraic BA approach for the XXZ open spin chain inspired by \cite{Cao2,BC}, the Sklyanin's separation of variables approach \cite{Niccoli} or the approach developed in \cite{Galleas} seem to be promising.

\section*{Acknowledgements}
  T.K. would like to thank Laboratoire de Math\'ematiques et Physique Th\'eorique, Universit\'e de Tours
for kind invitation and warm hospitality during his stay in March 2013.
This work is supported by the Grant-in-Aid for Scientific Research {\bf C} (21540228) from JSPS and Visiting professorship from CNRS.

\begin{appendix}

\section{Definitions of the $R$ and $K$ matrices}

\label{appendix:A}

In Sklyanin's framework \cite{Sklyanin},  the transfer matrix  associated with a finite spin chain is built from two objects: the  $R$-matrix and the $K-$matrix. For the model (\ref{def:Hamiltonian}), the $R$-matrix $R(\zeta)$ defined by:
\begin{eqnarray}
R(\zeta)=\frac{1}{\kappa(\zeta)}
\left(
\begin{array}{cccc}
1& & & \\
 &\frac{ (1-\zeta^2)q}{ 1-q^2\zeta^2}&\frac{
(1-q^2)\zeta}{
1-q^2\zeta^2}& \\
 &\frac{
(1-q^2)\zeta}{
1-q^2\zeta^2}&\frac{
(1-\zeta^2)q}{
1-q^2\zeta^2}& \\
 & & &1
\end{array}
\right),
\end{eqnarray}
where we have set
\begin{eqnarray}
\kappa(\zeta)=\zeta \frac{(q^4\zeta^2;q^4)_\infty (q^2/\zeta^2;q^4)_\infty}{
(q^4/\zeta^2;q^4)_\infty (q^2\zeta^2;q^4)_\infty},~~~
(z;p)_\infty=\prod_{n=0}^\infty (1-p^nz).\label{eq:kap}
\end{eqnarray}
As an operator on $V \otimes V$,
the matrix elements of $R(\zeta) \in {\rm End}(V \otimes V)$ are given by
$R(\zeta)v_{\epsilon_1}\otimes v_{\epsilon_2}=
\sum_{\epsilon_1', \epsilon_2'=\pm} 
v_{\epsilon_1'}\otimes v_{\epsilon_2'}
R(\zeta)_{\epsilon_1' \epsilon_2'}^{\epsilon_1 \epsilon_2}$,
where the ordering of the index is given by
$v_+\otimes v_+, v_+\otimes v_-, v_-\otimes v_+, v_-\otimes v_-$.
As usual, when copies $V_j$ of $V$ are involved, $R_{i j}(\zeta)$ acts as
$R(\zeta)$ on the $i$-th and $j$-th components and as identity elsewhere.
By definition, the $R$-matrix $R(\zeta)$ satisfies the Yang-Baxter equation :
\begin{eqnarray}
R_{1 2}(\zeta_1/\zeta_2)
R_{1 3}(\zeta_1/\zeta_3)
R_{2 3}(\zeta_2/\zeta_3)
=
R_{2 3}(\zeta_2/\zeta_3)
R_{1 3}(\zeta_1/\zeta_3)
R_{1 2}(\zeta_1/\zeta_2).
\end{eqnarray}
The normalization factor $\kappa(\zeta)$ (\ref{eq:kap}) is determined by the following unitarity and crossing symmetry conditions:
\begin{eqnarray}
R_{12}(\zeta)R_{21}(\zeta^{-1})=1,~~~
R(\zeta)_{\epsilon_2 \epsilon_1'}^{\epsilon_2' \epsilon_1}=
R(-q^{-1}\zeta^{-1})_{-\epsilon_1 \epsilon_2}^{-\epsilon_1' \epsilon_2'}.
\end{eqnarray}
The second basic object is the triangular $K$-matrix: 
$K^{(\pm)}(\zeta)=K^{(\pm)}(\zeta;r,s)$ \cite{dVG, GZ} by
\begin{eqnarray}
K^{(+)}(\zeta;r,s)
&=&
\frac{\varphi(\zeta^2;r)}{\varphi(\zeta^{-2};r)}
\left(\begin{array}{cc}
\frac{ 1-r\zeta^2}{ \zeta^2-r}&
\frac{ s \zeta(\zeta^2-\zeta^{-2})}{
 \zeta^2-r}\\
0&1
\end{array}
\right),
\label{def:K+}\\
K^{(-)}(\zeta;r,s)
&=&
\frac{\varphi(\zeta^2;r)}{\varphi(\zeta^{-2};r)}
\left(\begin{array}{cc}
\frac{
1-r\zeta^2}{
\zeta^2-r}&
0\\
\frac{
s \zeta (\zeta^2-\zeta^{-2})}{
\zeta^2-r}&1
\end{array}
\right),
\label{def:K-}
\end{eqnarray}
where we have set
\begin{eqnarray}
\varphi(z;r)=
\frac{(q^4rz;q^4)_\infty 
(q^6z^2;q^8)_\infty}{
(q^2 rz;q^4)_\infty 
(q^8z^2;q^8)_\infty}.
\end{eqnarray}
When viewed as an operator on $V$,
the matrix elements of $K^{(\pm)}(\zeta) \in {\rm End}(V)$ 
are given by
$K^{(\pm)}(\zeta)v_{\epsilon}=
\sum_{\epsilon'=\pm} 
v_{\epsilon'}
K^{(\pm)}(\zeta)_{\epsilon'}^{\epsilon}$,
where the ordering of the index is given by $v_+, v_-$.
As usual, when copies $V_j$ of $V$ are involved, $K_{j}^{(\pm)}(\zeta)$ acts as
$K^{(\pm)}(\zeta)$ on the $j$-th component and as identity elsewhere.
The $K$-matrix $K^{(\pm)}(\zeta)$ satisfies the boundary Yang-Baxter equation (also called the reflection equation):
\begin{eqnarray}
K_2^{(\pm)}(\zeta_2)
R_{21}(\zeta_1\zeta_2)
K_1^{(\pm)}(\zeta_1)
R_{12}(\zeta_1/\zeta_2)
=R_{21}(\zeta_1/\zeta_2)
K_1^{(\pm)}(\zeta_1)
R_{12}(\zeta_1\zeta_2)K_2^{(\pm)}(\zeta_2).
\label{eqn:BYBE}
\end{eqnarray}
The normalization factor (2.7) is determined by the following boundary unitarity and boundary crossing symmetry \cite{GZ}:
\begin{eqnarray}
K^{(\pm)}(\zeta)K^{(\pm)}(\zeta^{-1})=1,~~~
{K^{(\pm)}}(-q^{-1}\zeta^{-1})_{\epsilon_1}^{\epsilon_2}=
\sum_{\epsilon_1', \epsilon_2'=\pm}
R(-q\zeta^2)_{\epsilon_1' -\epsilon_2'}^{-\epsilon_1 \epsilon_2}
{K^{(\pm)}}(\zeta)_{\epsilon_2'}^{\epsilon_1'}.
\label{eqn:BUC}
\end{eqnarray}
The $K^{(\pm)}(\zeta;r,s)$ defined in (\ref{def:K+}) and (\ref{def:K-})
give general scalar triangular solutions of (\ref{eqn:BYBE}) and (\ref{eqn:BUC}). For $s=0$, one recovers the scalar diagonal solution of the reflection equation.

\section{Bosonizations of vertex operators}
\label{appendix:B}

In this Appendix we review
the bosonizations of the quantum group $U_q(\widehat{sl_2})$ and the $q$-vertex operators for level $k=1$ \cite{DFJMN, FJ}.
The center $\gamma^{\frac{1}{2}}$ of $U_q(\widehat{sl_2})$ satisfies $(\gamma^{\frac{1}{2}})^2=q^{h_0+h_1}=q^k$
on the level $k$ representation.
Hence $\gamma=q$ on the level $k=1$ representation.
For the level $k=1$ case,
the defining relation of $a_m~(m \in {\bf Z}_{\neq 0})$ becomes
\begin{eqnarray}
[a_m,a_n]=\delta_{m+n,0}\frac{[2m]_q[m]_q}{m}~~~(m,n\neq 0).
\end{eqnarray}
We introduce the zero-mode operator $\partial, \alpha$ by $[\partial,\alpha]=2$ and
the normal ordering symbol $:~:$
\begin{eqnarray}
:a_m a_n :=\left\{\begin{array}{cc}
a_m a_n &~(m<0)\\
a_n a_m &~(m>0)
\end{array}\right.,
~~~:\alpha \partial:=:\partial \alpha:=\alpha \partial.\label{def:normal-ordering}
\end{eqnarray}
The bosonizations of the irreducible highest representation $V(\Lambda_i)$ and its dual $V^*(\Lambda_i)$
with fundamental weight $\Lambda_i$
are given by 
\begin{eqnarray}
V(\Lambda_i)&=&{\bf C}[a_{-1},a_{-2},a_{-3}\cdots]\otimes \left(\oplus_{n \in {\bf Z}}{\bf C} e^{\Lambda_i+n \alpha}\right),\\
V^*(\Lambda_i)&=&{\bf C}[a_1,a_2,a_3,\cdots] \otimes \left(\oplus_{n \in {\bf Z}} {\bf C}e^{-\Lambda_i-n \alpha}\right).
\end{eqnarray}
On this space the actions of $a_m$, $e^\alpha$, $\partial$
are given by
\begin{eqnarray}
&& 
a_m (f \otimes e^\beta)=\left\{
\begin{array}{cc}
~a_m f \otimes e^\beta &~(m<0),\\
~[a_m,f] \otimes e^\beta &~(m>0),
\end{array}
\right.
\label{def:bosonization-action-1}\\
&&
e^\alpha (f \otimes e^\beta)=f \otimes e^{\alpha+\beta},~~
\partial (f \otimes e^\beta)=(\alpha, \beta)f \otimes e^\beta,
\label{def:bosonization-action-2}
\end{eqnarray}
where $f \in {\bf C}[a_{-1}, a_{-2}, \cdots]$, and
we have set $[\partial, \Lambda_0]=0$ and $\Lambda_1=\Lambda_0+\frac{\alpha}{2}$.

The action of other Drinfeld generators is given by
\begin{eqnarray}
&&K=q^{\partial},~~\gamma=q,\\
&&X^\pm(z)=\exp\left(R^\pm(z)\right)\exp\left(S^\pm(z)\right)e^{\pm \alpha}z^{\pm \partial},\\
&&q^d (1 \otimes e^{\Lambda_i+n \alpha})=q^{-(\beta,\beta)/2+i/4}(1\otimes e^{\Lambda_i+n \alpha}),
\end{eqnarray}
where 
\begin{eqnarray}
R^\pm(z)=\pm \sum_{n=1}^\infty \frac{a_{-n}}{[n]_q}q^{\mp n/2}z^n,~~~
S^\pm(z)=\mp\sum_{n=1}^\infty \frac{a_n}{[n]_q}q^{\mp n/2}z^{-n}.
\end{eqnarray}
%The relations between Chevalley generators and Drinfeld generators for level $c=1$ are given by
%\begin{eqnarray}
%h_1=\partial,~~~h_0=1-h_1,~~~
%e_1=x_0^+,~~~f_1=x_0^-,~~~
%e_0=x_1^- q^{-h_1},~~~f_0=q^{h_1}x_{-1}^+.
%\end{eqnarray}
We have the following bosonizations of the $q$-vertex operators :
\begin{eqnarray}
\Phi_-^{(1-i,i)}(\zeta)&=&\exp\left(P(\zeta^2)\right)\exp\left(Q(\zeta^2)\right)e^{\alpha/2}
(-q^3\zeta^2)^{(\partial+i)/2}\zeta^{-i},
\label{boson:VO1}
\\
\Phi_+^{(1-i,i)}(\zeta)&=&
\oint_{\widetilde{C}_1} \frac{dw}{2\pi\sqrt{-1}}\frac{(1-q^2)w\zeta}{
q(w-q^2\zeta^2)(w-q^4\zeta^2)}:\Phi_-^{(1-i,i)}(\zeta)X^-(w):,
\label{boson:VO2},
\end{eqnarray}
\begin{eqnarray}
\Psi_-^{* (1-i,i)}(\xi)&=&\exp\left(-P(q^{-1}\xi^2)\right)
\exp\left(-Q(q\xi^2)\right)e^{-\alpha/2}(-q^3\xi^2)^{(-\partial+i)/2}\xi^{1-i},\\
\Psi_+^{* (1-i,i)}(\xi)&=&
\oint_{\widetilde{C}_2}\frac{dw}{2\pi\sqrt{-1}}
\frac{q^2(1-q^2)\xi}{(w-q^2\xi^2)(w-q^4\xi^2)}:\Psi_-^{* (1-i,i)}(\xi)X^+(w):,
\end{eqnarray}
where
\begin{eqnarray}
P(z)=\sum_{n=1}^\infty
\frac{a_{-n}}{[2n]_q}q^{7n/2}z^n,~~~ Q(z)=-\sum_{n=1}^\infty \frac{a_n}{[2n]_q}q^{-5n/2}z^{-n}.
\end{eqnarray}
Here the integration contour $\widetilde{C}_1$ is a simple closed curve such that
the $w$ encircles $q^{4}\zeta^2$ inside but not $q^2\zeta^2$.
The integration contour $\widetilde{C}_2$ is a simple closed curve such that
the $w$ encircles $q^{2}\zeta^2$ inside but not $q^4 \zeta^2$.

In what follows we summarize normal orderings.
\begin{eqnarray}
&&
\Phi_-^{(i,1-i)}(\zeta_1)\Phi_-^{(1-i,i)}(\zeta_2)=
(-q^3 \zeta_1^2)^{\frac{1}{2}}
\frac{(q^2 \zeta_2^2/\zeta_1^2 ; q^4)_\infty}{
(q^4 \zeta_2^2/\zeta_1^2; q^4)_\infty}
:\Phi_-^{(i,1-i)}(\zeta_1)\Phi_-^{(1-i,i)}(\zeta_2):,
\\
&&
\Psi_-^{* (i,1-i)}(\xi_1)\Psi_-^{* (1-i,i)}(\xi_2)=
(-q^3 \xi_1^2)^{\frac{1}{2}}
\frac{(\xi_2^2/\xi_1^2;q^4)_\infty}{
(q^2 \xi_2^2/\xi_1^2;q^4)_\infty}:
\Psi_-^{* (1-i,i)}(\xi_1)\Psi_-^{* (1-i,i)}(\xi_2)
:,
\\
&&
\Phi_-^{(i,1-i)}(\zeta)\Psi_-^{* (1-i,i)}(\xi)=
(-q^3 \zeta^2)^{-\frac{1}{2}}
\frac{(q^3\xi^2/\zeta^2;q^4)_\infty}{
(q\xi^2/\zeta^2;q^4)_\infty}
:\Phi_-^{(i,1-i)}(\zeta)\Psi_-^{* (1-i,i)}(\xi):,
\\
&&\Psi_-^{* (i,1-i)}(\xi)\Phi_-^{(1-i,i)}(\zeta)=
(-q^3 \xi^2)^{-\frac{1}{2}}\frac{
(q^3 \zeta^2/\xi^2;q^4)_\infty}{(q \zeta^2/\xi^2;q^4)_\infty}
:
\Psi_-^{* (i,1-i)}(\xi)\Phi_-^{(1-i,i)}(\zeta):,
\\
&&X^+(w_1)X^+(w_2)=w_1^2 (1-w_2/w_1)(1-w_2/q^2 w_1):
X^+(w_1)X^+(w_2):,
\\
&&X^-(w_1)X^-(w_2)=w_1^2 (1-w_2/w_1)(1-q^2 w_2/w_1):
X^-(w_1)X^-(w_2)
:,
\\
&&X^+(w_1)X^-(w_2)=\frac{1}{w_1^2 (1-q w_2/w_1)(1-w_2/q w_1)}
:
X^+(w_1)X^-(w_2)
:,
\\
&&X^-(w_1)X^+(w_2)=\frac{1}{w_1^2 (1-q w_2/w_1)(1-w_2/q w_1)}
:
X^-(w_1)X^+(w_2)
:,
\\
&&\Phi_-^{(i,1-i)}(\zeta)X^+(w)=(-q^3 \zeta^2)(1-w/q^3\zeta^2)
:
\Phi_-^{(i,1-i)}(\zeta)X^+(w)
:,
\\
&&X^+(w)\Phi_-^{(i,1-i)}(\zeta)=w(1-q^3 \zeta^2/w)
:
X^+(w)\Phi_-^{(i,1-i)}(\zeta)
:,
\\
&&\Phi_-^{(i,1-i)}(\zeta)X^-(w)=\frac{-1}{q^3 \zeta^2 (1-w/q^2 \zeta^2)}
:
\Phi_-^{(i,1-i)}(\zeta)X^-(w)
:,
\\
&&X^-(w)\Phi_-^{(i,1-i)}(\zeta)=
\frac{1}{w(1-q^4 \zeta^2/w)}
:
X^-(w)\Phi_-^{(i,1-i)}(\zeta)
:,
\\
&&
\Psi_-^{* (i,1-i)}(\xi)X^+(w)
=\frac{-1}{q^3 \xi^2 (1-w/q^4 \xi^2)}
:
\Psi_-^{* (i,1-i)}(\xi)X^+(w)
:,
\\
&&
X^+(w)\Psi_-^{* (i,1-i)}(\xi)=\frac{1}{w (1-q^2 \xi^2/w)}
:
X^+(w)\Psi_-^{* (i,1-i)}(\xi)
:,
\\
&&
\Psi_-^{* (i,1-i)}(\xi)X^-(w)=(-q^3 \xi^2)
(1-w/q^3 \xi^2)
:
\Psi_-^{* (i,1-i)}(\xi)X^-(w)
:,
\\
&&
X^-(w)\Psi_-^{* (i,1-i)}(\xi)=w (1-q^3 \xi^2/w)
:
X^-(w)\Psi_-^{* (i,1-i)}(\xi):.
\end{eqnarray}

\section{Proof of the spin-reversal property}
\label{appendix:C}

In this Appendix we give a direct proof of one of the 
simplest example of the spin-reversal property.
We would like to show
\begin{eqnarray}
F^{(1,2,0)}_{-,+}(-q^{-1}\zeta,\zeta;r)=
F^{(0,2,0)}_{+,-}(-q^{-1}\zeta,\zeta;1/r).
\end{eqnarray}
We have
\begin{eqnarray}
F^{(i,2,0)}_{+,-}(-q^{-1}\zeta,\zeta;r)=\frac{
(q^2;q^2)_\infty^3
\Theta_{q^4}(\zeta^4)}{(1-\zeta^4)}
\oint_{C^{(i)}_{+-}}\frac{dw}{2\pi i}
\frac{(1-r\zeta^2)(1-\zeta^2 w/q^2)}{\zeta^2
(1-rw/q^2)}\frac{\Theta_{q^4}(q^2w^2)}{
\Theta_{q^2}(\zeta^2 w)\Theta_{q^2}(w/\zeta^2)}.
\end{eqnarray}
Here $C_{+-}^{(0)}$ is a closed curve satisfying that
the $w$ encircles 
$q^{2+2s}\zeta^2,~q^{2+2s}/\zeta^2~(s=0,1,2,\cdots)$
but not
$q^{-2s}\zeta^2,~q^{-2s}/\zeta^2~(s=0,1,2,\cdots),~q^2/r$.
Here $C_{+-}^{(1)}$ is a closed curve such that
$w$ encircles $q^2/r$ in addition the same points $C^{(0)}_{+-}$ does.
Upon the specialization we have
\begin{eqnarray}
F^{(i,2,0)}_{+,-}(-q^{-1}\zeta,\zeta;r)=
(-)\frac{
(q^2;q^2)_\infty^3
\Theta_{q^4}(z^2)}{(1-\zeta^4)}
\oint_{C^{(i)}_{-+}}\frac{dw}{2\pi i}
\frac{(1-r\zeta^2)(1-\zeta^2w/q^2)}{\zeta^2
(1-rw/q^2)}\frac{\Theta_{q^4}(q^2w^2)}{
\Theta_{q^2}(\zeta^2 w)\Theta_{q^2}(w/\zeta^2)}.
\end{eqnarray}
Here $C_{-+}^{(0)}$ is closed curve satisfying that
the $w$ encircles 
$q^{4+2s}\zeta^2,~q^{4+2s}/\zeta^2~(s=0,1,2,\cdots)$
but not
$q^{2-2s}\zeta^2,~q^{2-2s}/\zeta^2~(s=0,1,2,\cdots),~q^2/r$.
Here $C_{-+}^{(1)}$ is closed curve satisfying that
the $w$ encircles $q^2/r$ in addition the same points $C^{(0)}_{-+}$ does.
We note that the integrand does not have a pole at $w=q^2/\zeta^2$.
We have already deformed $C_{-+}^{(i)}$ from the original definition
of the correlation functions at $w=q^2/\zeta^2$.
Let's study
\begin{eqnarray}
F^{(0,2,0)}_{+,-}(-q^{-1}\zeta,\zeta;r)-
F^{(1,2,0)}_{-,+}(-q^{-1}\zeta,\zeta;1/r).
\end{eqnarray}
Changing $w \to q^4/w$ in the second term, we have
\begin{eqnarray}
\frac{(q^2;q^2)_\infty^3
\Theta_{q^4}(\zeta^4)}{\zeta^2}
\oint_{C_{+-}^{(0)}}\frac{dw}{2\pi i}
\frac{\Theta_{q^4}(q^2w^2)}{\Theta_{q^2}(w\zeta^2)\Theta_{q^2}(w/\zeta^2)}.
\end{eqnarray}
Taking the residues, we have
\begin{eqnarray}
\oint_{C_{+-}^{(0)}}\frac{dw}{2\pi i}
\frac{\Theta_{q^4}(q^2w^2)}{\Theta_{q^2}(w\zeta^2)\Theta_{q^2}(w/\zeta^2)}
=\sum_{n=1}^\infty (-q^2)^n \left({\rm Res}_{w=\zeta^2}+
{\rm Res}_{w=1/\zeta^2}\right)
\frac{\Theta_{q^4}(q^2w^2)}{\Theta_{q^2}(w\zeta^2)\Theta_{q^2}(w/\zeta^2)}=0.
\end{eqnarray}
Here we have used
$\frac{\Theta_{q^4}(q^{2-4n}w^2)}{\Theta_{q^2}(q^{-2n}w\zeta^2)
\Theta_{q^2}(q^{-2n}w/\zeta^2)}=
(-q^2)^n
\frac{\Theta_{q^4}(q^2w^2)}{\Theta_{q^2}(w\zeta^2)\Theta_{q^2}(w/\zeta^2)}$.

\section{Convenient formulae}

\label{appendix:D}

In this Appendix we summarize convenient formulae for calculations of
vacuum expectation values.
\begin{eqnarray}
\exp\left(\sum_{n=1}^\infty
\frac{1}{(q-q^{-1})}\frac{q^{(c+2)n}}{n[2n]_q}y^n
\right)&=&
(q^{c+4}y;q^4)_\infty,\\
\exp\left(\sum_{n=1}^\infty
\frac{1}{(q-q^{-1})}\frac{[n]_q q^{(c+2)n}}{n[2n]_q^2}y^n
\right)&=&\frac{
(q^{c+5}y;q^4,q^4)_\infty}{(q^{c+7}y;q^4,q^4)_\infty},\\
\exp\left(\sum_{n=1}^\infty
\frac{1}{(q-q^{-1})}\frac{q^{(c+2)n}}{n[n]_q}y^n
\right)&=&
(q^{c+3}y;q^4)_\infty,\\
\exp\left(\sum_{n>0}\frac{[n]_q}{n[2n]_q}\frac{q^{cn}}{1-\alpha_n \gamma_n}
y^n\right)
&=&
\frac{(q^{c+3}y;q^4,q^4)_\infty}{
(q^{c+1}y;q^4,q^4)_\infty},\\
\exp\left(
\sum_{n>0}\frac{[2n]_q}{n[n]_q}\frac{q^{cn}}{1-\alpha_n \gamma_n}y^n
\right)&=&\frac{1}{(q^{c-1}y;q^2)_\infty}.
\end{eqnarray}
%%%%%%%%%%%%%%%%%%%%%%%%%%%%%%
%%%%%%%%%%%%%%%%%%%%%%%%%%%%%%
\begin{eqnarray}
\exp\left(\sum_{n=1}^\infty \frac{[n]_q}
{n(1-\alpha_n\gamma_n)}\delta_n^{(i)}q^{cn/2}y^n\right)&=&\left(
\frac{(q^{c+3}y^2;q^8,q^8)_\infty
(q^{c+1}y^2;q^8,q^8)_\infty}{
(q^{c-1}y^2;q^8,q^8)_\infty
(q^{c+5}y^2;q^8,q^8)_\infty}
\right)^{1/2}\nonumber\\
&\times&
\left(\frac{(q^{(c-3)/2+2i}r^{1-2i}y;q^4,q^4)_\infty}{
(q^{(c+1)/2+2i}r^{1-2i}y;q^4,q^4)_\infty}\right)^{1-2i},\\
%%%%%%%%%%%%%%%%%%%%%%%%%%%%%%%%%%%%%%%%%%%
\exp\left(\sum_{n=1}^\infty \frac{[2n]_q}
{n(1-\alpha_n\gamma_n)}\delta_n^{(i)}q^{cn/2}y^n\right)&=&
\left(\frac{(q^{c-1}y^2;q^8)_\infty}{(q^{c-3}y^2;q^8)_\infty}\right)^{1/2}
(q^{(c-5)/2+2i} r^{1-2i}y;q^4)_\infty^{1-2i},\\
\exp\left(\sum_{n=1}^\infty \frac{[n]_q}
{n(1-\alpha_n\gamma_n)}
\beta_n^{(i)}q^{cn/2}y^n\right)&=&\left(
\frac{(q^{c+7}y^2;q^8,q^8)_\infty
(q^{c+13}y^2;q^8,q^8)_\infty}{
(q^{c+9}y^2;q^8,q^8)_\infty
(q^{c+11}y^2;q^8,q^8)_\infty}
\right)^{1/2}\nonumber\\
&\times&
\left(\frac{(q^{(c+9)/2-2i}r^{1-2i}y;q^4,q^4)_\infty}{
(q^{(c+13)/2-2i}r^{1-2i}y;q^4,q^4)_\infty}\right)^{1-2i},\\
\exp\left(\sum_{n=1}^\infty \frac{[2n]_q}
{n(1-\alpha_n\gamma_n)}\beta_n^{(i)}q^{cn/2}y^n\right)&=&
\left(\frac{(q^{c+5}y^2;q^8)_\infty}{(q^{c+7}y^2;q^8)_\infty}\right)^{1/2}
(q^{(c+7)/2-2i} r^{1-2i}y;q^4)_\infty^{1-2i}.
\end{eqnarray}
Here we have used $\alpha_n$, $\gamma_n$, $\beta_n^{(i)}$,
and $\delta_n^{(i)}$ in 
(\ref{def:alpha}), (\ref{def:beta}), and (\ref{def:delta}).
%%%%%%%%%%%%%%%%%%%%%%%%%%%%%%%%%

\end{appendix}

\end{document}